\UseRawInputEncoding

\documentclass[%
 reprint,
 amsmath,amssymb,
 aps,
]{revtex4-2}

\RequirePackage[T1]{fontenc}
\usepackage[utf8]{inputenc}
\usepackage{lmodern}

\usepackage{graphicx}
\usepackage{dcolumn}
\usepackage{bm}
\usepackage{float}
\usepackage{braket}
\usepackage{graphicx}
\usepackage{lipsum,capt-of,graphicx}
\usepackage{hyperref}
\hypersetup{
    colorlinks=true,
    linkcolor=blue,
    filecolor=magenta,   
    citecolor=blue,
    urlcolor=cyan,
    pdftitle={Overleaf Example},
    pdfpagemode=FullScreen,
    }
\usepackage[mathlines]{lineno}

\begin{document}

\preprint{APS/123-QED}

\title{Electron readout contrast enhancement in the parallel nuclear regime of an exchange-coupled donor spin qubit system}

\author{Holly G. Stemp$^{1,2}$}
    \altaffiliation[Currently at ]{Research Laboratory of Electronics, Massachusetts Institute of Technology, Cambridge, MA, USA}
\author{Mark R. van Blankenstein$^{1,2}$}
\author{Benjamin Wilhelm$^{1,2}$}
\author{Serwan Asaad$^{1,2}$}%
    \altaffiliation[Currently at ]{Quantum Machines, Tel Aviv, Israel}
\author{Mateusz T. M\k{a}dzik$^{1,2}$}%
    \altaffiliation[Currently at ]{Intel Corporation Hillsboro, Oregon, United States}
\author{Arne Laucht$^{1,2,3}$}%
\author{Fay E. Hudson$^{1,3}$}%
\author{Andrew S. Dzurak$^{1,3}$}%
\author{Kohei M. Itoh$^{4}$}%
\author{Alexander M. Jakob$^{2,5}$}%
\author{Brett C. Johnson$^{6}$}%
\author{David N. Jamieson$^{2,5}$}%
\author{Andrea Morello$^{1,2}$}%
 \email{a.morello@unsw.edu.au}

\affiliation{%
 $^{1}$ School of Electrical Engineering and Telecommunications, UNSW Sydney, Sydney, NSW 2052, Australia\\
 $^{2}$ ARC Centre of Excellence for Quantum Computation and Communication Technology\\
 $^{3}$ Diraq Pty. Ltd., Sydney, New South Wales, Australia\\
 $^{4}$ School of Fundamental Science and Technology, Keio University, Kohoku-ku, Yokohama, Japan \\
 $^{5}$ School of Physics, University of Melbourne, Melbourne, VIC 3010, Australia\\
 $^{6}$ School of Science, RMIT University, Melbourne, VIC, 3000, Australia
}%

\date{\today}
\begin{abstract}

Recent experiments on donor-based spin qubits in silicon have leveraged the exchange interaction between electrons bound to separate donor nuclei to perform two-qubit operations. A consistently observed yet unexplained phenomenon in such systems is the significant increase in electron readout contrast, measured via Elzerman-style readout to a single-electron transistor (SET) island, when the donor nuclei are initialized in a parallel spin orientation compared to an anti-parallel orientation. In this work, we present a detailed analysis of the exchange-coupled donor system in the parallel nuclear regime and propose a physical mechanism for this effect. We attribute the enhanced readout contrast to an additional electron tunneling event to the SET during a single read period, when the donor nuclei are aligned in a parallel spin configuration. These insights inform strategies for improving electron readout fidelity in these systems and contribute to a more complete understanding of spin-dependent tunnelling processes in donor-based qubit architectures.
\end{abstract}

\maketitle

\section{Introduction}\label{Introduction}
Donor spins in silicon are among the most promising qubit candidates for large-scale quantum information processing, offering exceptionally long coherence times, exceeding 30 seconds \cite{muhonen2014storing, chatterjee2021semiconductor} and single- and two-qubit gates with fidelities exceeding 99$\%$ \cite{dehollain2016optimization, pla2013high, stemp2024tomography, mkadzik2022precision}. Two-qubit interactions in this platform can be mediated in several ways. One approach couples two donor nuclei via a shared electron that is simultaneously hyperfine-coupled to both nuclear spins \cite{mkadzik2022precision, thorvaldson2025grover, edlbauer202511}. Another relies on the exchange coupling between electrons bound to separate donor atoms  \cite{kalra2014robust, madzik2021conditional, stemp2024tomography}. This exchange-based mechanism is both robust and scalable, enabling high-fidelity two-qubit controlled-rotation (CROT) gates between the electrons \cite{madzik2021conditional, stemp2024tomography}, as well as controlled-Z (CZ) operations between the nuclei via electron-mediated geometric phases \cite{stemp2025scalable}.\\

Owing to their central role in multi-qubit operations, exchange-coupled donor pairs are an essential building block for any scalable donor-based quantum processor. It is therefore crucial to fully characterize and understand their behavior. One intriguing feature consistently observed in exchange-coupled donor devices is a substantial increase in electron readout contrast when the donor nuclei are aligned in a parallel spin configuration ($\ket{\Downarrow \Downarrow}$ or $\ket{\Uparrow \Uparrow}$), compared to an anti-parallel configuration ($\ket{\Downarrow \Uparrow}$ or $\ket{\Uparrow \Downarrow}$), during Elzerman-style electron spin readout via spin-dependent tunneling to a single-electron-transistor (SET) reservoir \cite{morello2010single}. Across multiple devices, the readout contrast has been observed to increase by approximately 40$\%$ (for example, from 60$\%$ to 84$\%$) when the nuclei are initialized in a parallel rather than an anti-parallel spin orientation \cite{madzik2021conditional, stemp2024tomography}. Until now, the physical origin of this effect has not been studied.\\

In this work, we describe the physical mechanisms underlying Elzerman-style electron spin readout in exchange-coupled donor systems and explain the origin of this enhanced readout contrast for parallel nuclear configurations \cite{elzerman2004single, morello2010single}. We show that the enhancement arises from an additional tunneling event from the donor to the SET island when the nuclei are in a parallel spin orientation. We experimentally validate this explanation using a pair of exchange-coupled  $^{31}$P donors.\\

The remainder of this paper is organized as follows. In Section \ref{Description}, we describe the electron eigenstates of an exchange-coupled donor pair, for both parallel and antiparallel configurations of the donor nuclear spins. In Section \ref{Contrast}, we introduce a model that accounts for the enhanced electron readout contrast in the parallel spin configuration. Finally, in Section \ref{Results}, we present experimental evidence supporting this readout enhancement model.

\begin{figure*}[ht]
    \centering
    \includegraphics[width=0.9\textwidth]{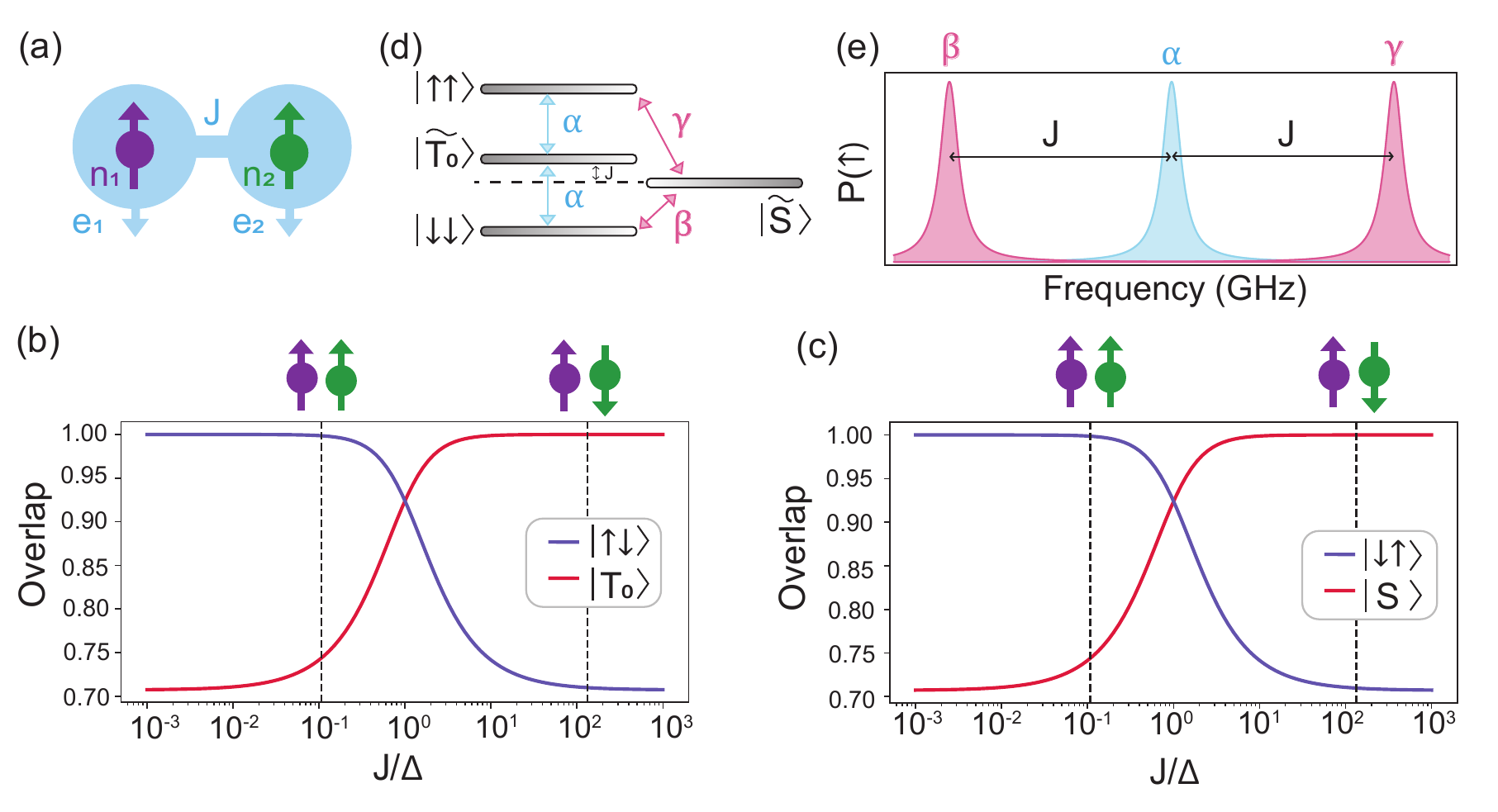}
   \caption[Description of the exchange-coupled donor system.]{Eigenstates of the exchange-coupled donor system. (a) Schematic of the exchange-coupled donor system, consisting of two donor nuclei: n$_{1}$ and n$_{2}$, each possessing a bound electron, e$_{1}$ and e$_{2}$, respectively. These electrons are coupled together with an exchange-interaction, $J$. (b) Projection of one of the electron eigenstates of the exchange-coupled system onto the pure state $\ket{\uparrow \downarrow}$ and onto the $\ket{T_0}$ state, as a function of the ratio between the electron coupling strength, $J$, and the electron detuning, $|\Delta|$. The black dashed lines show the typical $J/|\Delta|$ values for the case of the nuclei in either a parallel (10$^{-1}$) or anti-parallel (10$^{2}$) spin orientation. (c) Projection of the other electron eigenstates of the exchange-coupled system onto the pure state $\ket{\downarrow \uparrow }$ and onto the $\ket{S}$ state as a function of the ratio between the electron coupling strength, $J$, and the electron detuning, $|\Delta|$. (d) Energy eigenstates of the electrons in the parallel nuclei regime, where $J/|\Delta| \approx 10^{-1}$ . $\ket{T_{-}} = \ket{\downarrow \downarrow}$ and $\ket{T_{+}} = \ket{\uparrow \uparrow}$ represent two of the triplet states, while $\widetilde{\ket{S}}$ and $\widetilde{\ket{T_{0}}}$ represent a hybridised `singlet-like' or `triplet-like' state. (e) Schematic of the expected ESR transitions present in the parallel nuclear regime.}
    \label{fig:figure1}
\end{figure*}

\section{Description of exchange-coupled donor spins}\label{Description}

The exchange-coupled donor system considered in this work consists of a pair of $^{31}$P donor atoms, each hosting a single bound electron. In the present experiment, the donors are incorporated into the silicon lattice through ion implantation \cite{jakob2022deterministic, jamieson2013single}. Similar physics can be explored using donors introduced by scanning tunneling microscope (STM) lithography \cite{fuechsle2012single, schofield2025roadmap}. When the donors are placed sufficiently close to one another (typically within $\approx$ 20 nm) the spatial wavefunctions of the bound electrons overlap, giving rise to an exchange interaction, $J$, whose magnitude is highly sensitive to the precise inter-donor separation \cite{li2010exchange, burkard1999coupled,koiller2001exchange,klymenko2014electronic,saraiva2015theory,joecker2021full}. The system is typically operated in an external magnetic field, $B_0$, of order 1 T, which Zeeman-splits the electron (nuclear) spin states $\ket{\downarrow}(\ket{\Downarrow})$ and $\ket{\uparrow}(\ket{\Uparrow})$. Within a single donor, quantum information can be encoded in both the electron \cite{pla2012single} and the nuclear \cite{pla2013high} spin degrees of freedom, and the two spin qubit types can be efficiently entangled \cite{dehollain2016bell}.\\

The Hamiltonian of the two donor system, in the presence of a magnetic field along the Z-axis, is given by:

\begin{align}
        H = &(\mu_{\text{B}}/h) B_{0}(g_{1}S_{z1} + g_{2}S_{z2})+\\
    & \gamma_\mathrm{n}B_{0}(I_{z1}+I_{z2})+ \nonumber \\
    &  A_{1}\mathbf{S_{1}}\cdot \mathbf{I_{1}} +  A_{2}\mathbf{S_{2}}\cdot \mathbf{I_{2}} + \nonumber \\
    & J(\mathbf{S_{1}\cdot S_{2}}), \nonumber
\end{align}

\noindent where $\mu_{\text{B}}$ is the Bohr magneton, $h$ is the Planck's constant, $g_{1,2}\approx 1.9985$ the Land\'e  g-factors of each electron spin, $g\mu_{\text{B}}/h \approx$ 27.97 GHz/T and $\gamma_\mathrm{n} \approx$ -17.23 MHz/T is the $^{31}$P nuclear gyromagnetic ratio. $\mathbf{S_{1,2}}$ and $\mathbf{I_{1,2}}$ represent the vector spin operators for each electron and nucleus, and $A_{1,2}$ denotes the hyperfine couplings between the electron and nucleus for each donor atom.\\

In an exchange coupled donor system, the eigenstates of the electrons depend on the orientation of the donor nuclei. The electron eigenstates in a two-donor system, as depicted in the schematic in Fig. \ref{fig:figure1} (a), are given by:

\begin{align*}
    &\ket{\downarrow_1 \downarrow_2}\\
    &\widetilde{\ket{\downarrow_1 \uparrow_2}} = \cos(\theta)\ket{\downarrow_1 \uparrow_2} - \sin(\theta)\ket{\uparrow_1 \downarrow_2}\\
    &\widetilde{\ket{\uparrow_1 \downarrow_2 }} = \cos(\theta)\ket{\uparrow_1 \downarrow_2} + \sin(\theta)\ket{\downarrow_1\uparrow_2 }\\
    &\ket{\uparrow_1 \uparrow_2}
\end{align*}

\noindent where $\tan(2\theta) = \frac{J}{|\Delta|}$ and $|\Delta|$ represents the detuning between the electrons \cite{kalra2014robust, huang2019fidelity}.\\

The electron eigenstates, therefore, depend upon the ratio of the exchange interaction and the frequency detuning between the electrons, $\frac{J}{|\Delta|}$. In the limit $\frac{J}{|\Delta|} \ll 1$,  $\cos(\theta) \rightarrow 1$ and $\sin(\theta) \rightarrow 0$ and hence the electron eigenstates become the product states of the two electron states: $\ket{\downarrow_1 \downarrow_2}, \ket{\downarrow_1 \uparrow_2}, \ket{\uparrow_1 \downarrow_2}, \ket{\uparrow_1 \uparrow_2}$. On the other hand, in the limit $\frac{J}{|\Delta|} \gg 1$, $\cos(\theta) \rightarrow \frac{1}{\sqrt{2}}$ and $\sin(\theta) \rightarrow \frac{1}{\sqrt{2}}$. In this case, the electron eigenstates consist of the singlet and triplet states: $\ket{S} = \frac{1}{\sqrt{2}}(\ket{\downarrow_1 \uparrow_2} - \ket{\uparrow_1 \downarrow_2}), \ket{T_{-}} = \ket{\downarrow_1 \downarrow_2},  \ket{T_{0}} = \frac{1}{\sqrt{2}}(\ket{\downarrow_1 \uparrow_2} + \ket{\uparrow_1 \downarrow_2}), \ket{T_{+}} = \ket{\uparrow_1 \uparrow_2}$ \cite{kalra2014robust}. Figures \ref{fig:figure1} (b),(c) show the projection of the electron eigenstates on the $\ket{S}$ and $\ket{T_0}$ states and on the tensor product states $\ket{\downarrow \uparrow}$ and $\ket{\uparrow \downarrow}$ as a function of the ratio $\frac{J}{|\Delta|}$.\\

 In the donor spin system, we can transition between the two regimes of $\frac{J}{|\Delta|} \ll 1$, $\frac{J}{|\Delta|} \gg 1$ simply by preparing the two nuclei of the exchange-coupled system in either an anti-parallel or parallel spin orientation \cite{kalra2014robust}, provided that $J \ll A$. When initializing the two nuclei in an anti-parallel spin orientation ($\ket{\Downarrow_1 \Uparrow_2}$ or $\ket{\Uparrow_1 \Downarrow_2}$) the detuning between the two electrons is given by the average hyperfine interaction of the two donors $|\Delta| = \bar{A} = (A_{1} + A_{2})/2$. Typically, for $^{31}$P donor atoms in silicon, $\bar{A} \approx 117$ MHz \cite{pla2012single, feher1959electron}. Choosing an exchange interaction $\approx$ 10 MHz, as found in donors spaced by $\approx$ 20 nm \cite{joecker2021full}, places us in the regime $\frac{J}{|\Delta|} \approx$ 0.1 when the nuclear spins are in an anti-parallel state.\\

Conversely, for the case of the nuclei initialized in a parallel nuclear orientation, ($\ket{\Downarrow_1 \Downarrow_2}$ or $\ket{\Uparrow_1 \Uparrow_2}$), the detuning is given instead by the difference in hyperfine values between the two donors, $|\Delta| = \Delta A = |A_{1} - A_{2}|$. This may be caused by local variations of lattice strain \cite{usman2015strain,mansir2018linear,pla2018strain} and electric fields \cite{rahman2007high,pica2014hyperfine,laucht2015electrically}, and is typically of order a few MHz or less. In the system used in this work we measured $\Delta A = 90$ kHz, resulting in a $\frac{J}{|\Delta|} = $ 133. At this ratio of coupling to detuning, the two-electron odd-parity eigenstates deviate only slightly from perfect $\ket{S}$ and $\ket{T_0}$ states and hence we refer to these states as $\widetilde{\ket{S}}$ and $\widetilde{\ket{T_0}}$. The dashed vertical lines in Figs. \ref{fig:figure1} (b),(c) show the $\frac{J}{|\Delta|}$ values for the case of the nuclei in an anti-parallel and parallel orientation respectively. \\

Figures \ref{fig:figure1} (d),(e) show the electron spin resonance frequencies present when the nuclei are initialized in a parallel spin orientation. These consist of three transitions:

\begin{align}
    f_{\beta} &: \ket{\downarrow_1 \downarrow_2} \longleftrightarrow \widetilde{\ket{S}},\\
    f_{\alpha} &: \ket{\downarrow_1 \downarrow_2} \longleftrightarrow \widetilde{\ket{T_{0}}}, \widetilde{\ket{T_{0}}} \longleftrightarrow \ket{\uparrow_1 \uparrow_2},\\
    f_{\gamma} &: \widetilde{\ket{S}} \longleftrightarrow \ket{\uparrow_1 \uparrow_2},
\end{align}

More details on the transitions present in the parallel nuclear system are given in Appendix C.\\

For the remainder of this work we will focus entirely on the regime of $\frac{J}{|\Delta|} \gg 1$, achieved by initializing the two nuclei of the exchange-coupled system in a parallel spin orientation.

\section{Readout contrast enhancement in the parallel nuclear spins configuration}\label{Contrast}
Electron spin readout in a donor qubit system is typically performed via spin-dependent tunneling between donor-bound electrons and a nearby single-electron transistor (SET) \cite{morello2010single}. By positioning the SET electrochemical potential between the Zeeman-split spin levels, only a spin $\ket{\uparrow}$ electron can tunnel from the donor to the SET. This tunneling event temporarily shifts the SET out of Coulomb blockade, until a spin $\ket{\downarrow}$ electron tunnels from the SET onto the donor. We call this transient increase in SET current a `blip' of current. The spin state is inferred by thresholding the SET current during a fixed readout window, and the resulting ability to distinguish $\ket{\uparrow}$ from $\ket{\downarrow}$ defines the readout contrast. Readout contrast is reduced by errors such as thermally activated tunneling and missed blips, the latter occurring when tunneling events are faster than the bandwidth of the readout electronics and therefore fail to cross the detection threshold. A detailed description of the readout mechanism and associated error processes is provided in Appendix A.\\

We have observed that, when measuring the electron spin of a donor pair where the nuclei are initialized in a parallel spin orientation, the electron spin readout contrast is increased compared to the case of the nuclei in an anti-parallel spin orientation. We hypothesize that this is due to the presence of an additional spin $\ket{\uparrow}$ electron, resulting in two tunneling events to the SET island during a single read period. This extra tunneling event presents an additional opportunity to identify a current blip, thus reducing the instances of missed current blips and the subsequent mis-identification of a spin $\ket{\uparrow}$ electron as a spin $\ket{\downarrow}$. \\

\begin{figure*}[ht]
    \centering
    \includegraphics[width=1\textwidth]{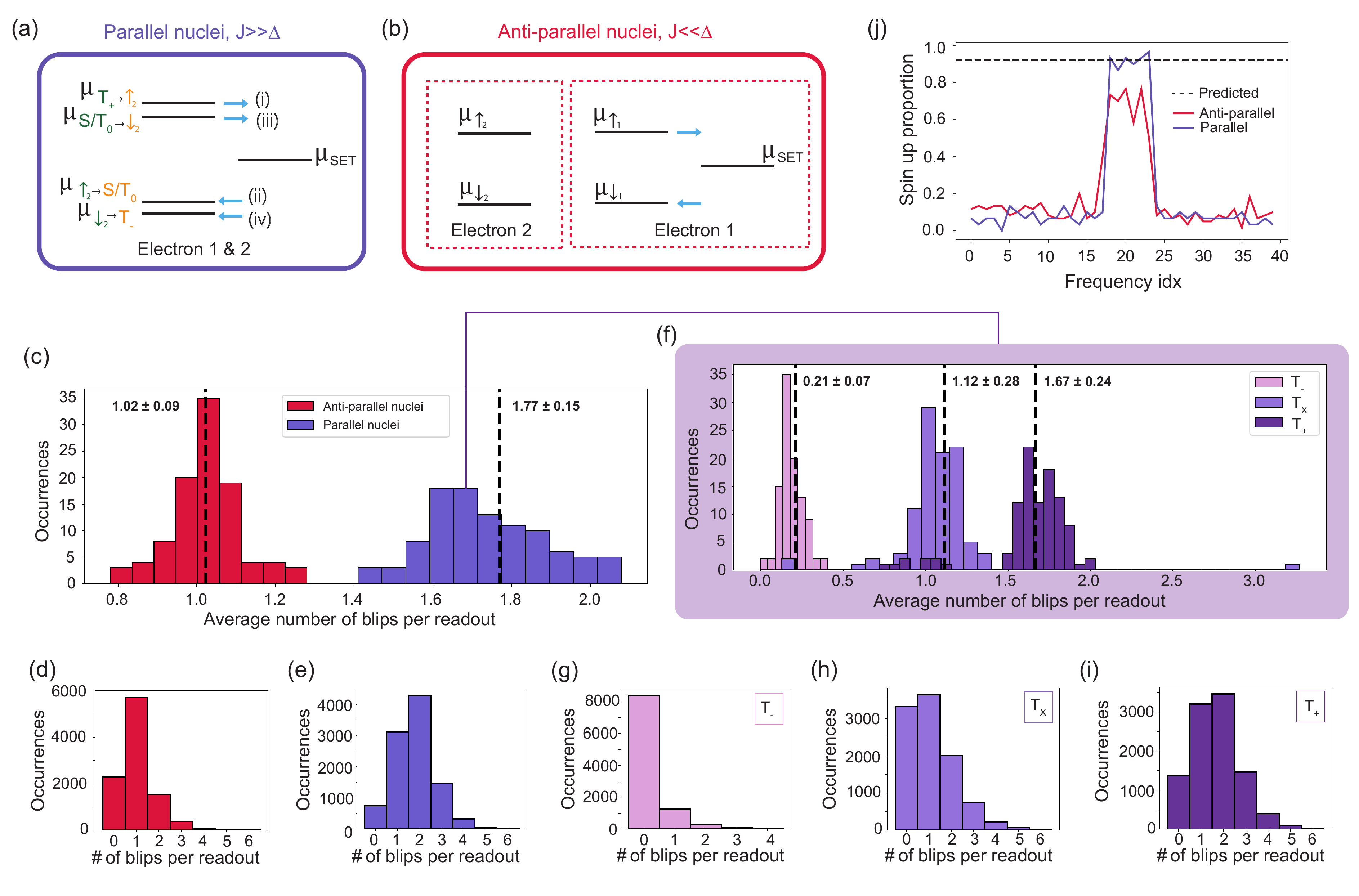}
   \caption[Tunneling to the SET in the parallel vs anti-parallel nuclear regime.]{Tunneling to the SET in the parallel vs anti-parallel nuclear regime. (a) Two-electron energy levels with respect to the electrochemical potential of the SET, $\mu_{\text{SET}}$. Labels (i - iv) depict the numbered steps associated with the electron readout process for the case of the nuclei in a parallel spin orientation. The subscripts for each of the two-electron energy levels depict the initial (green) and final (yellow) states following a tunneling event. (b) Tunneling scheme for the case of the anti-parallel nuclei. In this case the eigenstates of the two electrons are separable and hence the tunneling scheme of electron 1 is the same as for a single-donor electron. (c) Histogram of average number of blips per readout for the case of the nuclei being in a parallel (purple) or anti-parallel (red) state. The dashed lines represent the mean value of each histogram. The error bar for this value, and throughout this work, is given by one standard deviation, $1\sigma$. Histogram of the number of blips per readout for the case of the (d) anti-parallel or (e) parallel nuclei. (f) Histogram of the average number of blips per readout trace when initializing in either the $\ket{T_-}$, $\ket{T_{X}}$ or $\ket{T_+}$ state, where $\ket{T_{X}}$ is a mixture of $\widetilde{\ket{{T_0}}}$, $\ket{T_-}$ and $\ket{T_+}$ (see Appendix C). (g),(h),(i) Histogram of the number of blips per readout for the case of parallel nuclei, when the electrons are initialized in a (g) $\ket{T_-}$, (h) $\widetilde{\ket{{T_0}}}$ and (i) $\ket{T_+}$ state.  (j) Adiabatic frequency spectrum of one of the electrons in the exchange-coupled donor pair, for the case of the nuclei being initialized in either an anti-parallel spin orientation (red line) or a parallel spin orientation (purple line). The black dashed line shows the expected up proportion for the case of the parallel nuclear regime, given in equation \ref{up_proportion_formula}, where  $P_{(\Downarrow \Downarrow / \Uparrow \Uparrow)}$ and $P_{(\Downarrow \Uparrow/\Uparrow \Downarrow)}$ represent the electron spin up proportions for the parallel and anti-parallel nuclear states, respectively.}
    \label{fig:figure_2}
\end{figure*}

The process behind this double tunneling event is outlined in Fig. \ref{fig:figure_2} (a), where one electron of an exchange-coupled donor atom pair is read out via spin dependent tunneling to the SET. In the case of a parallel spin configuration of the two donor nuclei, since the eigenstates of the two electrons include the entangled $\widetilde{\ket{T_0}}$ and $\widetilde{\ket{S}}$ states, we must consider the combined, two-electron energy levels of the system in order to understand the tunneling behavior of electron 1 \cite{watson2015high}. The energy level diagram for the exchange-coupled donor pair system is shown in Fig. \ref{fig:figure_2} (a).\\

When the nuclei are in a parallel spin configuration, such that $J \gg |\Delta| = \Delta A$, the following dynamics are expected after initializing the electrons in the $\ket{T_+} = \ket{\uparrow_{1} \uparrow_{2}}$ state and moving to the readout position (neglecting erroneous tunneling events arising from finite temperature):

\begin{enumerate}
    \item A spin $\ket{\uparrow_{1}}$ electron tunnels from donor 1 to the SET island, producing a current blip and leaving behind a spin $\ket{\uparrow_{2}}$ electron on donor 2 [Fig. \ref{fig:figure_2}(a)-i]. 
    \item While remaining at the readout position, a spin $\ket{\downarrow_{1}}$ electron will eventually tunnel from the SET island back onto the ionized donor 1 [Fig. \ref{fig:figure_2} (a)-ii]. Since the electron must tunnel into an eigenstate of the system, and the other electron of the pair is in the $\ket{\uparrow_{2}}$ state, the most energetically favorable state is either the $\widetilde{\ket{T_0}}$ or $\widetilde{\ket{S}}$ state. These states are separated by $J \approx 10$ MHz which is negligible compared to both the electron Zeeman splitting, $\gamma_e B_0\approx$~28 GHz, and the thermal broadening $5k_{\rm B}T/h \approx 10$~GHz at 100~mK. As a result, the probability of tunneling into either the  the $\widetilde{\ket{T_0}}$ or $\widetilde{\ket{S}}$ state is approximately equal. Both states have odd parity, corresponding to a single spin excitation shared between the two electrons. In general, one can expect different tunnel rates into the $\widetilde{\ket{T_0}}$ and $\widetilde{\ket{S}}$ states, because they correspond to different orbital wavefunctions \cite{hanson2005single,dehollain2014single}, however, we expect this difference to be negligible for the low $J$ values used in this work.
    \item After waiting at the readout position for some proportion of the spin $\ket{\uparrow_1}$ tunnel time, a second spin $\ket{\uparrow_1}$ electron will tunnel from donor 1 to the SET island [Fig. \ref{fig:figure_2} (a)-iii]. This is because the electron tunnels from a $\widetilde{\ket{T_0}}$ or $\widetilde{\ket{S}}$ state, which both have a significant $\ket{\uparrow_1}$ probability. This tunneling event leaves behind a spin $\ket{\downarrow_{2}}$ electron on donor 2, as electron 2 must be left in the opposite spin state to the $\ket{\uparrow_1}$ electron. This tunneling event results in a second blip of current through the SET.

    \item Another $\ket{\downarrow_{1}}$ electron will tunnel from the SET island onto donor 1, leaving the electrons in the state $\ket{T_{-}} = \ket{\downarrow_{1}\downarrow_{2}}$ [Fig. \ref{fig:figure_2} (a)-iv]. With both electrons now in the $\ket{\downarrow}$ state, any further tunneling events are energetically forbidden for the remainder of the read period.
\end{enumerate}

For the case of the anti-parallel nuclei, in the energy regime where $J \ll |\Delta| = \bar{A}$, the eigenstates of the electrons are, to a good approximation, separable product states of electron 1 and electron 2 spin states. Therefore, when considering the tunneling of electron 1, the relevant energy levels are the single-electron levels associated with that electron, as shown in Fig. \ref{fig:figure_2} (b). In this case, the readout proceeds identically to the standard electron readout described in Appendix A.\\

Following initialization of a spin $\ket{\uparrow_1}$, the readout dynamics are as follows:

\begin{enumerate}
    \item A spin $\ket{\uparrow_{1}}$ electron will tunnel off of donor 1, to the SET island, producing a blip of current.
    \item After waiting at the readout position for some proportion of the tunnel time, a spin $\ket{\downarrow_{1}}$ electron tunnels from the SET to donor 1, preventing any further tunneling during the remainder of the read period.
\end{enumerate}

\section{Results}\label{Results}

\subsection{Counting readout blips in the parallel vs anti-parallel nuclear configuration}

The experiments described in the following sections were carried out on a system of two $^{31}$P donor atoms, coupled by an exchange interaction $J \approx 10$ MHz and a difference in hyperfine coupling between the two donors atoms of $\Delta A \approx 90$ kHz. Only electron 1 was tunnel-coupled to the SET and hence could be read out via spin-dependent tunneling.  Fig \ref{fig:tunneling_times} in Appendix B shows the measured tunnel-in rate of the $\ket{\downarrow_{1}}$ electron 1 from the SET island to the donor in this device, which was measured to be 32.8 $\pm$ 0.4 $\mu$s. The error bars here and throughout this work indicate $1\sigma$. The bandwidth of our measurement setup can be estimated as $0.35/t_r$ where $t_{r}$ is the rise time of the SET current blip, defined as the time taken for the current to rise from 10 - 90$\%$ of its maximum value. Extracting the blip rise time directly from the measured SET current traces, we obtain $t_{r} =$ 6.7 $\pm$ 0.2 $\mu$s, corresponding to a measurement bandwidth of approximately 50 kHz, consistent with the settings of the transimpedance amplifier. We thus expect there to be a significant fraction of missed SET current blips in this device, due to the tunnel time approaching the limit of the instrument bandwidth. \\

We first reproduced the pulse sequences for which this increase in electron readout contrast had been observed in prior experiments. This is typically observed when applying a $\pi$-pulse to one of the electrons of the donor pair, with the nuclei initialized in either a parallel or anti-parallel orientation, followed by an electron readout.\\

For the case in which the nuclei were in an anti-parallel configuration, we initialized the two nuclear spins in the $\ket{\Uparrow_{1} \Downarrow_{2}}$ state and the electrons in the  $\ket{\downarrow_{1} \downarrow_{2}}$ state. We then applied a $\pi$-pulse to flip electron 1 conditional on electron 2 being in the $\ket{\downarrow_{2}}$ state, resulting in the preparation of the two-electron state $\widetilde{\ket{\uparrow_{1} \downarrow_{2}}}$. \\

For the parallel nuclear configuration, we initialized the nuclei in the $\ket{\Downarrow_{1} \Downarrow_{2}}$ state and again prepared the electrons in the  $\ket{\downarrow_{1} \downarrow_{2}}$ state. We then applied a $\pi$-pulse to electron 1 at the frequency corresponding to the $f_{\alpha}$ transition. Because the two electrons are almost degenerate in the parallel nuclear case, this operation drives both electrons and thus prepares the two electron state $\ket{T_{+}} = \ket{\uparrow_{1} \uparrow_{2}}$ (see Appendix C). \\

For both the anti-parallel and parallel nuclear case, after performing the $\pi$ pulse, we then read out the state of electron 1, by moving to the readout position in gate space and waiting there for 1 ms. The experiment was repeated 10,000 times, in order to obtain 10,000 SET readout current traces. For each current trace the number of current blips per readout trace was counted.\\

Figure \ref{fig:figure_2} (c) shows a histogram of the average number of current blips per SET readout trace, for the case of the parallel (purple) and anti-parallel (red) nuclear state. This was obtained by arbitrarily dividing the 10,000 total experiment repetitions into 100 subgroups, with 100 readout traces per subgroup. Counting how many blips were observed per readout trace and averaging over the 100 traces per subgroup, we are able to plot a histogram of the average blip number observed per readout. For the case of the anti-parallel nuclear state the average number of blips per current trace is centered at 1.02 $\pm$ 0.09, indicating that on average there is close to one tunneling event per readout period, as expected. For the case of the parallel nuclear state, the average number of blips per readout is instead centered at 1.77 $\pm$ 0.15, indicating that on average, more than one current blip occurs per readout trace. The reason that this value is lower than 2 is most likely due to missed current blips.\\

To better understand the distribution in the number of blips per readout period, it is also helpful to look at the raw number of blip events per readout, without any averaging. Figure \ref{fig:figure_2} (d),(e) shows histograms of the number of blips counted per readout trace for the 10,000 experiment repetitions, without any subgroup averaging, for the case of the nuclei being in an anti-parallel (panel d) or parallel (panel e) orientation. This reveals that, for the case of the nuclei being initialized in an anti-parallel spin orientation, the majority of readout traces consist of a single tunneling event. Conversely, for the case of the parallel nuclei, the majority of readout traces consist of two tunneling events. \\

\subsection{Counting blips in the parallel nuclear configuration}
To further investigate whether these increased SET blip events can be attributed to the tunneling process described in Fig.\ref{fig:figure_2} (a), we performed a more detailed analysis of each of the electron eigenstates when the nuclei are initialized in a parallel spin state. In this experiment, the nuclei were once again initialized in the parallel spin $\ket{\Downarrow_1 \Downarrow_2}$ state, however, this time the experiment was repeated for three separate initial electron states. For each initial state we performed 10,000 readouts of electron 1:

\begin{enumerate}
    \item \textbf{Electrons initialized in the $\ket{T_{-}} = \ket{\downarrow_1 \downarrow_2}$ state.} This was done using our standard electron initialization procedure, using spin dependent tunneling from the SET island. 
    \item \textbf{Electrons initialized in  in the $\ket{T_X}$ state.} Due to the $\ket{T_{-}} \leftrightarrow \widetilde{\ket{T_0}}$ and $\widetilde{\ket{T_0}} \leftrightarrow \ket{T_{+}}$ transitions being almost degenerate, the $\widetilde{\ket{T_0}}$ is difficult to prepare experimentally. Instead we prepare the $\ket{T_X}$ state, which consists of a mixture of the $\widetilde{\ket{T_0}}$, $\ket{\downarrow_1 \downarrow_2}$ and $\ket{\uparrow_1 \uparrow_2}$ states. This state was prepared by first initializing the electrons in the $\ket{T_{-}} = \ket{\downarrow_1 \downarrow_2}$ and then performing a $\frac{\pi}{2}$-pulse at the transition frequency $f_{\alpha}$ (see Appendix C for details).
    \item \textbf{Electrons initialized in the $\ket{T_{+}} = \ket{\uparrow_1 \uparrow_2}$ state.} This was done by once again initializing the electrons in the $\ket{\downarrow_1 \downarrow_2}$ state, followed by performing a $\pi$-pulse at a frequency $f_{\alpha}$.
\end{enumerate}

Figure \ref{fig:figure_2}(f) shows a histogram of the average number of blips (averaged over 100 subgroups as described previously) for the case of each of the three initial states outlined above. An initial state of $\ket{T_{-}}$ results in an average number of blips of 0.21 $\pm$ 0.07. Figure \ref{fig:figure_2}(g) shows a histogram of the number of blips per readout across the 10,000 readout traces, revealing that the majority of traces showed no readout blips. The traces that did show readout blips can likely be attributed to the errors associated with electron initialization and readout at finite temperature as described above.\\

An initial state $\ket{T_{X}}$ results in an average number of 1.12 $\pm$ 0.28 blips per readout. Figure \ref{fig:figure_2}(h) shows a histogram of the number of blips per readout for this initial state. As the $\ket{T_{X}}$ state consists of a mixture of $\approx$ 50$\%$ $\widetilde{\ket{T_0}}$, $\approx$ 25$\%$ $\ket{T_{-}}$ and $\approx$ 25$\%$ $\ket{T_{+}}$ we expect each measurement to project this state onto either the $\ket{T_{-}}$, $\widetilde{\ket{T_0}}$ or $\ket{T_{+}}$ state, which should result in either 0, 1 or 2 readout blips respectively. The histogram in Fig. \ref{fig:figure_2}(h) appears to show a large proportion of readout traces with either 0, 1 or 2 readout blips, which would indeed be consistent with the preparation of the $\ket{T_{X}}$ initial state.\\

Finally, an initial state $\ket{T_+}$ results in an average number of 1.67 $\pm$ 0.24 blips. Figure \ref{fig:figure_2} (i) shows the histogram of the number of blips per readout for this initial state, revealing that the majority of traces showed two readout blips. Many traces showed a single blip, again likely due to missing blip events. The instances where more than two blips were present during a single readout trace were also likely due to temperature induced readout errors (see Appendix A). For example, the erroneous tunneling of a spin $\ket{\uparrow}$ electron from the SET to the donor, which can result in an additional blip during the readout period. \\

\subsection{Electron spin up proportion}
As an additional test of our hypothesis regarding the origin of the increased electron readout contrast, we used this model to calculate the expected electron spin-up proportion when the nuclei are in the parallel configuration. Assuming that missed blips are the dominant mechanism reducing the observed electron spin-up proportion below 1 in the anti-parallel nuclear case, the probability of missing a blip can be approximated as $ 1-P_{(\Downarrow \Uparrow/\Uparrow \Downarrow)}$ where $P_{(\Downarrow \Uparrow/\Uparrow \Downarrow)}$ is the measured electron spin up proportion when the donor nuclei are anti-parallel. In this case, the probability of missing two current blips is given by  $(1-P_{(\Downarrow \Uparrow/\Uparrow \Downarrow)})^{2}$. The expected electron spin up proportion for the case of parallel nuclei would then be given by:

\begin{equation}
\label{up_proportion_formula}
    P_{(\Downarrow \Downarrow / \Uparrow \Uparrow)} = 1 - (1-P_{(\Downarrow \Uparrow/\Uparrow \Downarrow)})^{2}
\end{equation}

Figure \ref{fig:figure_2}(j) shows a spectroscopy experiment with an adiabatically chirped pulse, performed on one electron in the exchange-coupled donor pair, with the nuclei initialized in either an anti-parallel spin orientation (red line) or parallel spin orientation (purple line). The adiabatic pulse is insensitive to small frequency jumps \cite{laucht2014high}, which could cause our driving pulse to be off-resonance and thus reduce the measured spin up proportion. We can therefore assume that the height of the resonance peaks in an adiabatic frequency spectrum is dominated by missed readout blips and not control errors. $P_{(\Downarrow \Uparrow/\Uparrow \Downarrow)}$ is therefore extracted directly from the height of the resonance peak for the anti-parallel nuclei. The black dashed line in Fig. \ref{fig:figure_2}(j) shows the predicted spin up proportion for the parallel nuclei, given by equation \ref{up_proportion_formula}. This line has good correspondence with the experimentally measured height of the resonance peak for the parallel nuclei, further validating the theory that the increased readout contrast is caused by two SET current blips per readout trace being present for the parallel nuclei.

\section{Summary and Outlook}
We have described the principles governing electron spin readout in an exchange-coupled donor system when the donor nuclei are initialized in a parallel spin orientation. We have shown that the $\approx 40 \%$ enhancement in electron spin readout contrast observed in this regime originates from an additional tunneling event to the SET island within the single readout period. This additional tunneling event reduces the instances of missed current blips, that occur when the return tunneling of an electron from the SET to the donor is too rapid to be resolved by the measurement bandwidth. \\

Understanding this mechanism not only clarifies how readout contrast is enhanced in the parallel-nuclear case, but also provides broader insight into the dynamics of exchange-coupled donor systems. These insights can inform strategies to improve readout fidelity for both parallel and anti-parallel nuclear configurations, and contribute to a more complete understanding of spin-dependent tunneling processes in donor-based qubit architectures.

\section*{Acknowledgments}
This research was funded by the Australian Research Council Centre of Excellence for Quantum Computation and Communication Technology (CE170100012) and the US Army Research Office (Contracts no. W911NF-17-1-0200 and W911NF-23-1-0113). A.M. acknowledges an Australian Research Council Laureate Fellowship (FL240100181). We acknowledge the facilities, and the scientific and technical assistance provided by the UNSW node of the Australian National Fabrication Facility (ANFF), and the Heavy Ion Accelerators (HIA) nodes at the University of Melbourne and the Australian National University. ANFF and HIA are supported by the Australian Government through the National Collaborative Research Infrastructure Strategy (NCRIS) program. H.G.S., M.R.v.B. and B.W. acknowledge support from the Sydney Quantum Academy. The views and conclusions contained in this document are those of the authors and should not be interpreted as representing the official policies, either expressed or implied, of the Army Research Office or the U.S. Government. The U.S. Government is authorized to reproduce and distribute reprints for Government purposes notwithstanding any copyright notation herein.

\section*{Data availability}
The data that support the findings of this article are openly available at \cite{stemp2026paralleldata}.

\section*{Appendix A: Electron readout}\label{Readout}

In this section we review the readout method used for donor spin spin qubits using the typical architecture comprising a single-electron transistor and a tunnel-coupled donors \cite{morello2009architecture}. This method is well established and universally adopted across ion-implanted \cite{morello2010single,pla2012single,laucht2014high,madzik2021conditional,mkadzik2022precision} and STM-fabricated \cite{buch2013spin} donor devices.

\subsection*{Readout procedure}

The spin state of the electron in a donor qubit is read out via spin-dependent tunneling to a single-electron transistor (SET) reservoir \cite{elzerman2004single, morello2010single}. The SET island consists of a large quantum-dot containing $\approx$ 100 electrons. At this electron occupancy, the energy levels of the SET island form a semi-continuum of states \cite{morello2009architecture}, which can be described by a thermally broadened Fermi-Dirac distribution. In this scenario, the electrochemical potential of the SET island, $\mu_{\text{SET}}$, essentially coincides with a Fermi energy. \\

\begin{figure}[ht]
    \centering
    \includegraphics[width=0.5\textwidth]{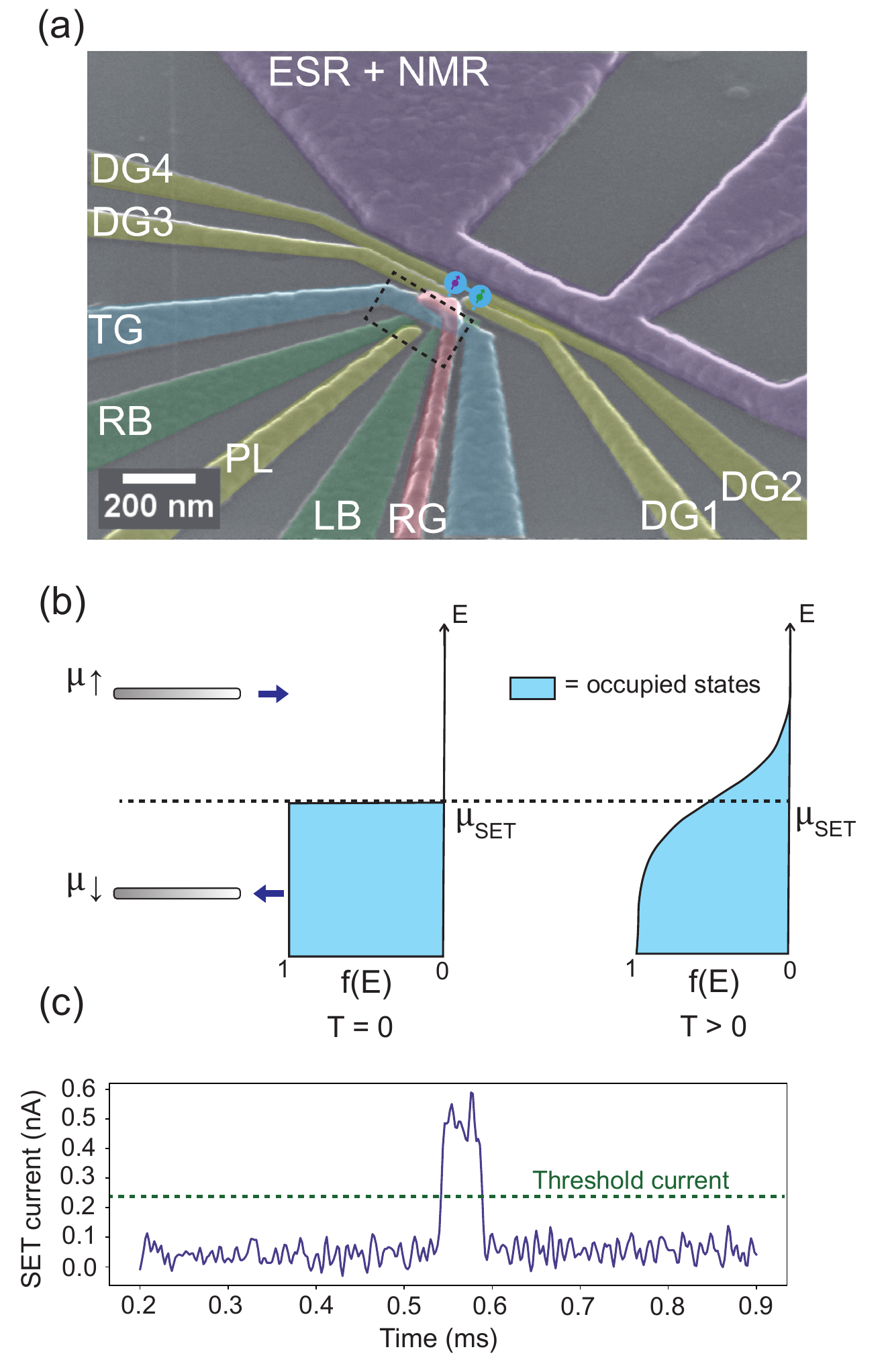}
    \caption[Elzerman-style readout of the donor electron spin.]{Elzerman-style readout of the donor electron spin. (a) False-colored scanning electron microscope image of the gate structure of a device nominally the same as the device used in this work. The donor atom schematics shows the location of the donor implantation window in the device. The single-electron transistor (SET) region is highlighted by the dashed black box. (b) Schematics of the Fermi-Dirac distributions that describe the distribution of states in the SET island for the case of a temperature of \textit{T} = 0 K (left) and temperature \textit{T} > 0 K (right). $\mu_{\text{SET}}$ represents the electrochemical potential of the SET island. (c) SET current trace during readout, showing a `blip' of current, indicating the tunneling of an $\ket{\uparrow}$ electron to the SET island and re-initialization into the $\ket{\downarrow}$ state. The dashed green line indicates an example of a threshold current, which is used to determine whether or not a blip occurred during a readout trace.}
    \label{fig:figure_readout}
\end{figure}

The devices are fabricated such that the SET island lies in close proximity (typically $\sim 20$~nm \cite{mohiyaddin2013noninvasive}) to the donor atoms, with the donor-bound electrons tunnel-coupled to the SET. Figure \ref{fig:figure_readout} (a) shows a false-colored scanning electron microscope image of a typical device, with the location of the donor atoms with respect to the SET island (black dashed box) highlighted. By manipulating the relative energy of $\mu_{\text{SET}}$ and the Zeeman split spin energy levels of the electron, $\mu_{\downarrow}, \mu_{\uparrow}$, using local gate voltages,  $\mu_{\text{SET}}$ can be positioned in-between $\mu_{\uparrow}$ and $\mu_{\downarrow}$, as depicted in Fig. \ref{fig:figure_readout} (b). As $\mu_{\uparrow}$ is now higher in energy than $\mu_{\text{SET}}$ while $\mu_{\downarrow}$ is now lower in energy than $\mu_{\text{SET}}$, only a spin $\ket{\uparrow}$ electron is facing unoccupied energy states in the SET island and thus can tunnel from the donor to the SET. This tunneling event leaves behind a positive charge at the site of the ionized donor nucleus, which shifts the SET island out of Coulomb blockade, resulting in a current flowing through the SET. Only a spin $\ket{\downarrow}$ electron has sufficient energy to tunnel from the SET island back onto the donor. Once this happens, the SET is once again shifted back into Coulomb blockade and the current through the SET is blocked. We call this increase in SET current for the duration of the tunnel-time for a $\ket{\downarrow}$ electron to tunnel back onto the donor a `blip' of current. \\

A spin $\ket{\uparrow}$ electron can be identified by setting a threshold current. If the SET current rises above this threshold during the readout window, then this indicates a tunneling event from the donor to the SET and hence the electron was in the $\ket{\uparrow}$ state. Figure \ref{fig:figure_readout} (c) shows an example SET current trace showing a `blip' event, corresponding to the tunneling of an $\ket{\uparrow}$ electron from the donor to the SET and initialization into the $\ket{\downarrow}$ state. Conversely, if the SET remains below this threshold current throughout the duration of the readout window, then this indicates that no tunneling event occurred and hence the electron was in the spin $\ket{\downarrow}$ state \cite{morello2010single, johnson2022beating}. \\

The ability to correctly distinguish the presence of a spin $\ket{\downarrow}$ electron from the presence of a spin $\ket{\uparrow}$ electron during the readout period is referred to as the readout contrast, which spans between 0 and 1.

\subsection*{Errors in electron readout}
One source of error associated with the readout of the donor electron can be attributed to thermal broadening of the Fermi–Dirac distribution in the SET island. Figure \ref{fig:figure_readout} (b) shows a schematic of the Fermi–Dirac distribution describing the available states in the SET island. At a temperature of \textit{T} = 0 K, the distribution is a perfect step function, with all states below $\mu_{\text{SET}}$ occupied and all states above $\mu_{\text{SET}}$ unoccupied. At finite temperature (\textit{T} > 0 K), this distribution broadens by $\approx 5k_{B}T$, which leads to readout errors due to thermally activated tunneling events \cite{morello2010single, johnson2022beating}.\\

A second source of readout error arises from missed current blips. Missed blips occur when, despite the donor electron being in the $\ket{\uparrow}$ state, the tunneling rate to the SET island is so fast that the SET current does not rise above the detection threshold during the readout window. As a result, the electron is incorrectly identified as being in the $\ket{\downarrow}$ state. When the tunneling time to/from donor and SET is much faster than the rise/fall time of the readout electronics, missed blips can become the dominant contribution to the total readout error. Although the incidence of missed blips can be reduced by increasing the bandwidth of the readout instrumentation, this typically comes at the expense of a lower signal-to-noise ratio in the detected current, which limits how large a bandwidth can be practically employed \cite{huang2021high}.\\

One additional source of error is the erroneous flipping of the nucleus following the readout of the electron, which we refer to as ionization shock \cite{joecker2024error, vaartjes2025maximizing}. This ionization shock comes about because of the weak hybridization between the nucleus and electron, which can cause the readout routine to deviate from a truly quantum non-demolition (QND) process. There is the potential that the increased number of blips in the parallel nuclear orientation could lead to an increase in ionization shock events. However, this was not studied explicitly in this work.

\section*{Appendix B: Measuring tunnel-in times}

\begin{figure}[ht]
    \centering
    \includegraphics[width=0.5\textwidth]{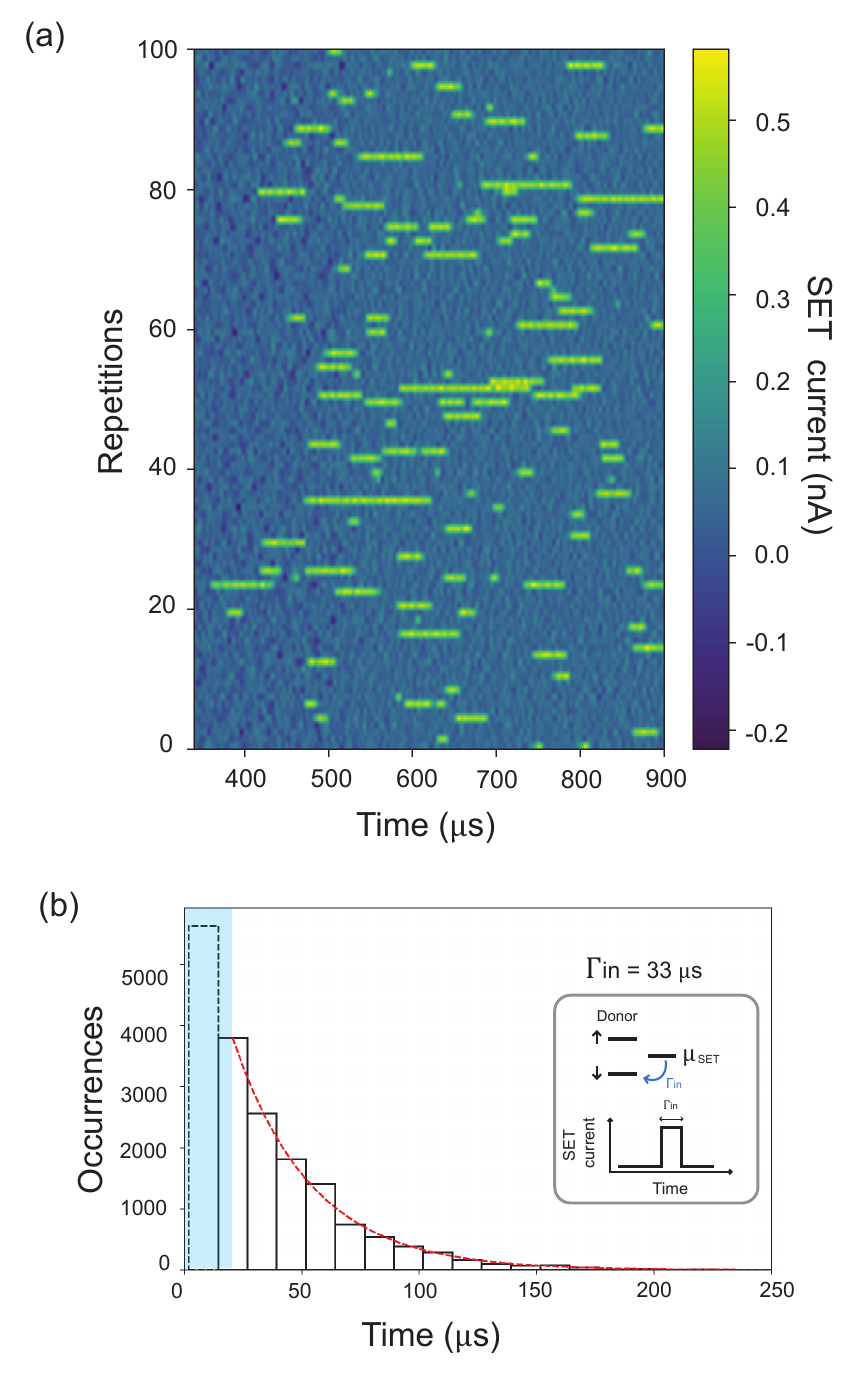}
    \caption[tunneling time calculation.]{Tunneling time calculation. (a) 100 raw SET current traces.  (b) Histogram of the tunneling time of the electron from the SET island to the donor taken from 100 measured SET current traces. The red line shows the exponential fit to this histogram, from which we extracted the tunnel time. The blue box shows the bandwidth of the instruments, which was approximately 50 kHz in these experiments. Durations below the bandwidth are not accurately determined and are denoted with a dashed box.}
    \label{fig:tunneling_times}
\end{figure}

The duration of a readout `blip' is determined by the typical tunnel-in time of a $\ket{\downarrow}$ electron, from SET island to donor. Figure \ref{fig:tunneling_times} (a) shows 100 raw SET traces taken from the dataset used to obtain the plots in Fig. \ref{fig:figure_2} of the main text. By calculating the blip duration in each of the SET traces, plotting this in a histogram and fitting this histogram to an exponential decay (Fig. \ref{fig:tunneling_times} (b)), we obtain a tunnel-in time of 32.8 $\pm$ 0.4 $\mu$s. Here, we are excluding the first histogram bar from the fit as blip durations shorter than the measurement bandwidth are not accurately determined.

\section*{Appendix C: Operation of the electrons in the parallel nuclear regime}
\label{appendix_C}

\begin{figure}[ht]
    \centering
    \includegraphics[width=0.5\textwidth]{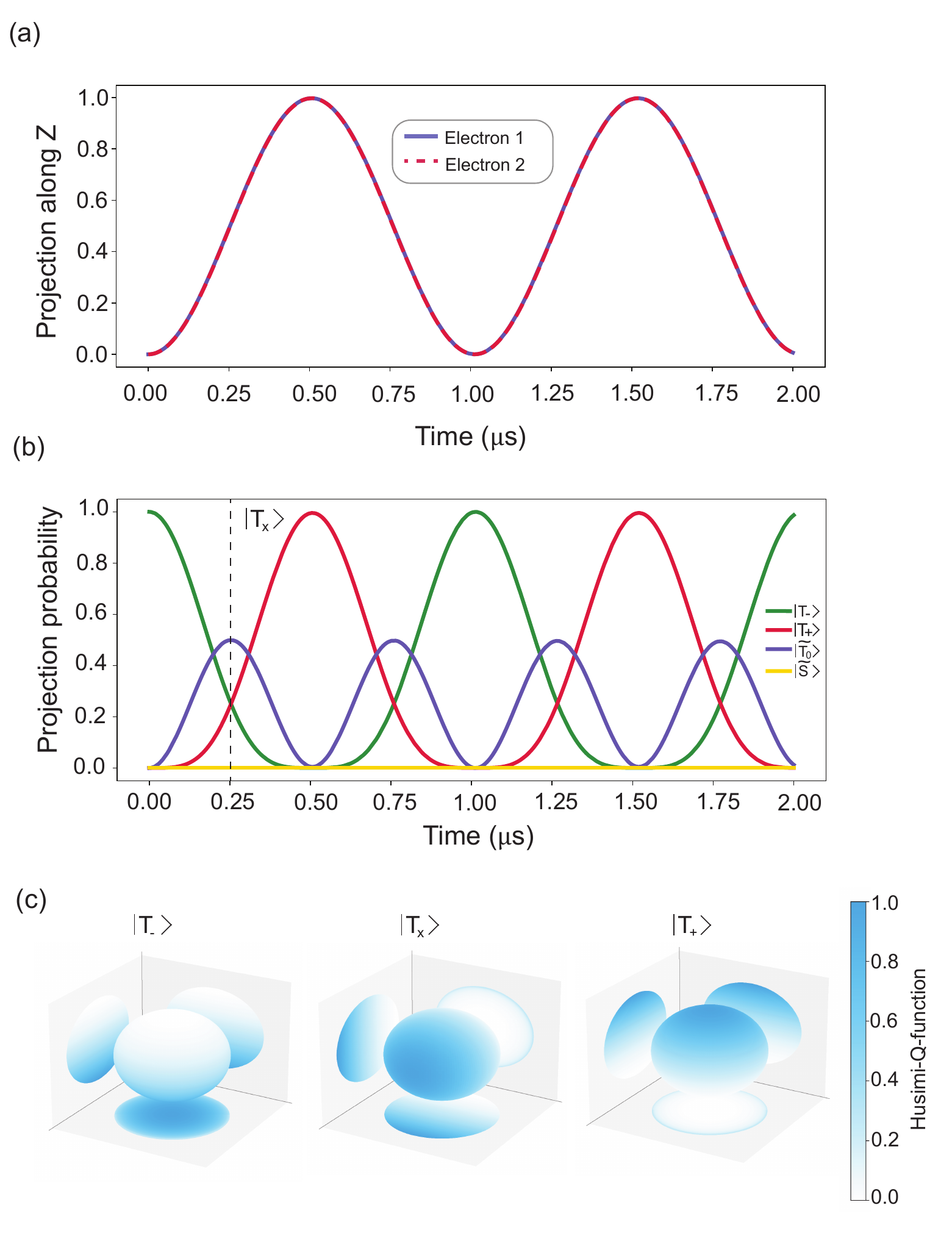}
    \caption[Electron spin time evolution simulation in the parallel nuclear regime]{Electron spin time evolution simulation in the parallel nuclear regime. (a) Simulation of the driving of the $\alpha$ transition, at a frequency centered between the $\ket{T_-}$ $\leftrightarrow$ $\ket{T_0}$ and $\ket{T_0}$ $\leftrightarrow$ $\ket{T_+}$ transition, which differ due to a different hyperfine between the two donors. The simulation was carried out with $J$ = 10 MHz and $\Delta A$ = 90 kHz and a Rabi frequency of $f_R = $ 1 MHz. The electrons were initialized in the $T_- = \ket{\downarrow_1 \downarrow_2}$ state before applying the driving pulse. This plot shows the projection of the two electrons along the Z-axis during the time evolution. (b) The same time evolution shown in (a) but with the projection on each of the two-electron eigenstates plotted over time. Vertical black dashed line denotes the $\frac{\pi}{2}$ duration at which the state $\ket{T_{X}}$ is prepared. (c) Husimi functions of the $\ket{T_-}$,$\ket{T_{X}}$ and $\ket{T_{+}}$ states. }
    \label{fig:parallel_sim}
\end{figure}

When the nuclei of a exchange-coupled donor system are initialized in a parallel spin orientation, the following electron spin resonance (ESR) transitions are present:

\begin{align}
    f_{\beta} &: \ket{\downarrow_1 \downarrow_2} \longleftrightarrow \widetilde{\ket{S}},\\
    f_{\alpha} &: \ket{\downarrow_1 \downarrow_2} \longleftrightarrow \widetilde{\ket{T_{0}}}, \widetilde{\ket{T_{0}}} \longleftrightarrow \ket{\uparrow_1 \uparrow_2},\\
    f_{\gamma} &: \widetilde{\ket{S}} \longleftrightarrow \ket{\uparrow_1 \uparrow_2},
\end{align}

For the case in which $|\Delta| A = 0$, the transitions $\ket{\downarrow_1 \downarrow_2} \longleftrightarrow \widetilde{\ket{T_{0}}}, \widetilde{\ket{T_{0}}} \longleftrightarrow \ket{\uparrow_1 \uparrow_2}$ are degenerate and thus a driving tone applied at a frequency $f_{\alpha}$ will result in a covariant $SU(2)$ rotation within the manifold of triplet states with total spin $S=1$ \cite{yu2025schrodinger}. For the case in which $|\Delta| A \neq 0$ these transitions become split by approximately $\sqrt{J^{2}+|\Delta| A^2}-J$. However, provided $2f_{R} > |\Delta| A$, where $f_{R} $ is the Rabi frequency of the electron, the linewidth of the applied driving field is larger than the frequency splitting between the transitions and hence both transitions will be driven simultaneously -- this is known as the `hard pulse' regime in nuclear magnetic resonance. In this work we use $f_{R} \approx 1$ MHz and $\sqrt{J^{2}+|\Delta| A^2}-J \approx 140$ Hz and hence a driving field applied at a frequency $f_{\alpha}$ will result in a simultaneous driving of the $\ket{\downarrow_1 \downarrow_2} \longleftrightarrow \widetilde{\ket{T_{0}}}$ and $\widetilde{\ket{T_{0}}} \longleftrightarrow \ket{\uparrow_1 \uparrow_2}$ transitions.\\

The transitions $\ket{\downarrow_1 \downarrow_2} \longleftrightarrow \widetilde{\ket{S}}$ and $\widetilde{\ket{S}} \longleftrightarrow \ket{\uparrow_1 \uparrow_2}$ are present due to $\widetilde{\ket{S}}$ not representing a pure singlet and hence this state possessing a small but nonzero spin, which can be addressed with ESR \cite{kalra2014robust, nakai1991forbidden}.\\

Figures \ref{fig:parallel_sim}(a)-(b) show the simulated time evolution of the electrons upon applying an oscillating magnetic drive centred between the $\ket{T_-}$ $\leftrightarrow$ $\widetilde{\ket{T_0}}$ transition and $\widetilde{\ket{T_0}}$ $\leftrightarrow$ $\ket{T_+}$ transition. A value of $J$ = 10 MHz, $\Delta A = 90$ kHz and Rabi frequency of $f_{R} = $ 1 MHz was used in these simulations. Figure \ref{fig:parallel_sim} (a) shows the projection of this time evolution along the Z-axis of the Bloch sphere, revealing that both electrons undergo Rabi oscillations as a result of the applied drive rotating both electrons simultaneously. Figure \ref{fig:parallel_sim} (b) shows the projection of the time evolved state on the eigenstates $\ket{\downarrow_1 \downarrow_2}, \widetilde{\ket{T_0}},  \widetilde{\ket{S}}$ and $\ket{\uparrow_1 \uparrow_2}$. Applying a pulse at a frequency $f_{\alpha}$, with a duration of $\pi$ results in the electrons undergoing a rotation between the $\ket{\downarrow_1 \downarrow_2}$ and $\ket{\uparrow_1 \uparrow_2}$ state. Moreover, applying a pulse at $f_{\alpha}$ for a $\frac{\pi}{2}$ duration (as denoted by the vertical dashed line in Fig. \ref{fig:parallel_sim} (b)) results in a transition between the $\ket{\downarrow_1 \downarrow_2}$ state and a state made up of a mixture of the $\widetilde{\ket{T_0}}, \ket{\downarrow_1 \downarrow_2}$ and $\ket{\uparrow_1 \uparrow_2}$, which we denote the $\ket{T_{X}}$ state. Such state is a coherent spin state pointing along the equator of the $S=1$ Bloch sphere of the triplet states, as shown in the Husimi function in Fig. \ref{fig:parallel_sim} (c).\\


\begin{thebibliography}{47}%
\makeatletter
\providecommand \@ifxundefined [1]{%
 \@ifx{#1\undefined}
}%
\providecommand \@ifnum [1]{%
 \ifnum #1\expandafter \@firstoftwo
 \else \expandafter \@secondoftwo
 \fi
}%
\providecommand \@ifx [1]{%
 \ifx #1\expandafter \@firstoftwo
 \else \expandafter \@secondoftwo
 \fi
}%
\providecommand \natexlab [1]{#1}%
\providecommand \enquote  [1]{``#1''}%
\providecommand \bibnamefont  [1]{#1}%
\providecommand \bibfnamefont [1]{#1}%
\providecommand \citenamefont [1]{#1}%
\providecommand \href@noop [0]{\@secondoftwo}%
\providecommand \href [0]{\begingroup \@sanitize@url \@href}%
\providecommand \@href[1]{\@@startlink{#1}\@@href}%
\providecommand \@@href[1]{\endgroup#1\@@endlink}%
\providecommand \@sanitize@url [0]{\catcode `\\12\catcode `\$12\catcode `\&12\catcode `\#12\catcode `\^12\catcode `\_12\catcode `\%12\relax}%
\providecommand \@@startlink[1]{}%
\providecommand \@@endlink[0]{}%
\providecommand \url  [0]{\begingroup\@sanitize@url \@url }%
\providecommand \@url [1]{\endgroup\@href {#1}{\urlprefix }}%
\providecommand \urlprefix  [0]{URL }%
\providecommand \Eprint [0]{\href }%
\providecommand \doibase [0]{https://doi.org/}%
\providecommand \selectlanguage [0]{\@gobble}%
\providecommand \bibinfo  [0]{\@secondoftwo}%
\providecommand \bibfield  [0]{\@secondoftwo}%
\providecommand \translation [1]{[#1]}%
\providecommand \BibitemOpen [0]{}%
\providecommand \bibitemStop [0]{}%
\providecommand \bibitemNoStop [0]{.\EOS\space}%
\providecommand \EOS [0]{\spacefactor3000\relax}%
\providecommand \BibitemShut  [1]{\csname bibitem#1\endcsname}%
\let\auto@bib@innerbib\@empty
\bibitem [{\citenamefont {Muhonen}\ \emph {et~al.}(2014)\citenamefont {Muhonen}, \citenamefont {Dehollain}, \citenamefont {Laucht}, \citenamefont {Hudson}, \citenamefont {Kalra}, \citenamefont {Sekiguchi}, \citenamefont {Itoh}, \citenamefont {Jamieson}, \citenamefont {McCallum}, \citenamefont {Dzurak} \emph {et~al.}}]{muhonen2014storing}%
  \BibitemOpen
  \bibfield  {author} {\bibinfo {author} {\bibfnamefont {J.~T.}\ \bibnamefont {Muhonen}}, \bibinfo {author} {\bibfnamefont {J.~P.}\ \bibnamefont {Dehollain}}, \bibinfo {author} {\bibfnamefont {A.}~\bibnamefont {Laucht}}, \bibinfo {author} {\bibfnamefont {F.~E.}\ \bibnamefont {Hudson}}, \bibinfo {author} {\bibfnamefont {R.}~\bibnamefont {Kalra}}, \bibinfo {author} {\bibfnamefont {T.}~\bibnamefont {Sekiguchi}}, \bibinfo {author} {\bibfnamefont {K.~M.}\ \bibnamefont {Itoh}}, \bibinfo {author} {\bibfnamefont {D.~N.}\ \bibnamefont {Jamieson}}, \bibinfo {author} {\bibfnamefont {J.~C.}\ \bibnamefont {McCallum}}, \bibinfo {author} {\bibfnamefont {A.~S.}\ \bibnamefont {Dzurak}}, \emph {et~al.},\ }\bibfield  {title} {\bibinfo {title} {Storing quantum information for 30 seconds in a nanoelectronic device},\ }\href@noop {} {\bibfield  {journal} {\bibinfo  {journal} {Nature Nanotechnology}\ }\textbf {\bibinfo {volume} {9}},\ \bibinfo {pages} {986} (\bibinfo {year} {2014})}\BibitemShut {NoStop}%
\bibitem [{\citenamefont {Chatterjee}\ \emph {et~al.}(2021)\citenamefont {Chatterjee}, \citenamefont {Stevenson}, \citenamefont {De~Franceschi}, \citenamefont {Morello}, \citenamefont {de~Leon},\ and\ \citenamefont {Kuemmeth}}]{chatterjee2021semiconductor}%
  \BibitemOpen
  \bibfield  {author} {\bibinfo {author} {\bibfnamefont {A.}~\bibnamefont {Chatterjee}}, \bibinfo {author} {\bibfnamefont {P.}~\bibnamefont {Stevenson}}, \bibinfo {author} {\bibfnamefont {S.}~\bibnamefont {De~Franceschi}}, \bibinfo {author} {\bibfnamefont {A.}~\bibnamefont {Morello}}, \bibinfo {author} {\bibfnamefont {N.~P.}\ \bibnamefont {de~Leon}},\ and\ \bibinfo {author} {\bibfnamefont {F.}~\bibnamefont {Kuemmeth}},\ }\bibfield  {title} {\bibinfo {title} {Semiconductor qubits in practice},\ }\href@noop {} {\bibfield  {journal} {\bibinfo  {journal} {Nature Reviews Physics}\ }\textbf {\bibinfo {volume} {3}},\ \bibinfo {pages} {157} (\bibinfo {year} {2021})}\BibitemShut {NoStop}%
\bibitem [{\citenamefont {Dehollain}\ \emph {et~al.}(2016{\natexlab{a}})\citenamefont {Dehollain}, \citenamefont {Muhonen}, \citenamefont {Blume-Kohout}, \citenamefont {Rudinger}, \citenamefont {Gamble}, \citenamefont {Nielsen}, \citenamefont {Laucht}, \citenamefont {Simmons}, \citenamefont {Kalra}, \citenamefont {Dzurak} \emph {et~al.}}]{dehollain2016optimization}%
  \BibitemOpen
  \bibfield  {author} {\bibinfo {author} {\bibfnamefont {J.~P.}\ \bibnamefont {Dehollain}}, \bibinfo {author} {\bibfnamefont {J.~T.}\ \bibnamefont {Muhonen}}, \bibinfo {author} {\bibfnamefont {R.}~\bibnamefont {Blume-Kohout}}, \bibinfo {author} {\bibfnamefont {K.~M.}\ \bibnamefont {Rudinger}}, \bibinfo {author} {\bibfnamefont {J.~K.}\ \bibnamefont {Gamble}}, \bibinfo {author} {\bibfnamefont {E.}~\bibnamefont {Nielsen}}, \bibinfo {author} {\bibfnamefont {A.}~\bibnamefont {Laucht}}, \bibinfo {author} {\bibfnamefont {S.}~\bibnamefont {Simmons}}, \bibinfo {author} {\bibfnamefont {R.}~\bibnamefont {Kalra}}, \bibinfo {author} {\bibfnamefont {A.~S.}\ \bibnamefont {Dzurak}}, \emph {et~al.},\ }\bibfield  {title} {\bibinfo {title} {Optimization of a solid-state electron spin qubit using gate set tomography},\ }\href@noop {} {\bibfield  {journal} {\bibinfo  {journal} {New Journal of Physics}\ }\textbf {\bibinfo {volume} {18}},\ \bibinfo {pages} {103018} (\bibinfo {year} {2016}{\natexlab{a}})}\BibitemShut {NoStop}%
\bibitem [{\citenamefont {Pla}\ \emph {et~al.}(2013)\citenamefont {Pla}, \citenamefont {Tan}, \citenamefont {Dehollain}, \citenamefont {Lim}, \citenamefont {Morton}, \citenamefont {Zwanenburg}, \citenamefont {Jamieson}, \citenamefont {Dzurak},\ and\ \citenamefont {Morello}}]{pla2013high}%
  \BibitemOpen
  \bibfield  {author} {\bibinfo {author} {\bibfnamefont {J.~J.}\ \bibnamefont {Pla}}, \bibinfo {author} {\bibfnamefont {K.~Y.}\ \bibnamefont {Tan}}, \bibinfo {author} {\bibfnamefont {J.~P.}\ \bibnamefont {Dehollain}}, \bibinfo {author} {\bibfnamefont {W.~H.}\ \bibnamefont {Lim}}, \bibinfo {author} {\bibfnamefont {J.~J.}\ \bibnamefont {Morton}}, \bibinfo {author} {\bibfnamefont {F.~A.}\ \bibnamefont {Zwanenburg}}, \bibinfo {author} {\bibfnamefont {D.~N.}\ \bibnamefont {Jamieson}}, \bibinfo {author} {\bibfnamefont {A.~S.}\ \bibnamefont {Dzurak}},\ and\ \bibinfo {author} {\bibfnamefont {A.}~\bibnamefont {Morello}},\ }\bibfield  {title} {\bibinfo {title} {High-fidelity readout and control of a nuclear spin qubit in silicon},\ }\href@noop {} {\bibfield  {journal} {\bibinfo  {journal} {Nature}\ }\textbf {\bibinfo {volume} {496}},\ \bibinfo {pages} {334} (\bibinfo {year} {2013})}\BibitemShut {NoStop}%
\bibitem [{\citenamefont {Stemp}\ \emph {et~al.}(2024)\citenamefont {Stemp}, \citenamefont {Asaad}, \citenamefont {Blankenstein}, \citenamefont {Vaartjes}, \citenamefont {Johnson}, \citenamefont {M{\k{a}}dzik}, \citenamefont {Heskes}, \citenamefont {Firgau}, \citenamefont {Su}, \citenamefont {Yang} \emph {et~al.}}]{stemp2024tomography}%
  \BibitemOpen
  \bibfield  {author} {\bibinfo {author} {\bibfnamefont {H.~G.}\ \bibnamefont {Stemp}}, \bibinfo {author} {\bibfnamefont {S.}~\bibnamefont {Asaad}}, \bibinfo {author} {\bibfnamefont {M.~R.~v.}\ \bibnamefont {Blankenstein}}, \bibinfo {author} {\bibfnamefont {A.}~\bibnamefont {Vaartjes}}, \bibinfo {author} {\bibfnamefont {M.~A.}\ \bibnamefont {Johnson}}, \bibinfo {author} {\bibfnamefont {M.~T.}\ \bibnamefont {M{\k{a}}dzik}}, \bibinfo {author} {\bibfnamefont {A.~J.}\ \bibnamefont {Heskes}}, \bibinfo {author} {\bibfnamefont {H.~R.}\ \bibnamefont {Firgau}}, \bibinfo {author} {\bibfnamefont {R.~Y.}\ \bibnamefont {Su}}, \bibinfo {author} {\bibfnamefont {C.~H.}\ \bibnamefont {Yang}}, \emph {et~al.},\ }\bibfield  {title} {\bibinfo {title} {Tomography of entangling two-qubit logic operations in exchange-coupled donor electron spin qubits},\ }\href@noop {} {\bibfield  {journal} {\bibinfo  {journal} {Nature Communications}\ }\textbf {\bibinfo {volume} {15}},\ \bibinfo {pages} {8415} (\bibinfo {year} {2024})}\BibitemShut
  {NoStop}%
\bibitem [{\citenamefont {M{\k{a}}dzik}\ \emph {et~al.}(2022)\citenamefont {M{\k{a}}dzik}, \citenamefont {Asaad}, \citenamefont {Youssry}, \citenamefont {Joecker}, \citenamefont {Rudinger}, \citenamefont {Nielsen}, \citenamefont {Young}, \citenamefont {Proctor}, \citenamefont {Baczewski}, \citenamefont {Laucht} \emph {et~al.}}]{mkadzik2022precision}%
  \BibitemOpen
  \bibfield  {author} {\bibinfo {author} {\bibfnamefont {M.~T.}\ \bibnamefont {M{\k{a}}dzik}}, \bibinfo {author} {\bibfnamefont {S.}~\bibnamefont {Asaad}}, \bibinfo {author} {\bibfnamefont {A.}~\bibnamefont {Youssry}}, \bibinfo {author} {\bibfnamefont {B.}~\bibnamefont {Joecker}}, \bibinfo {author} {\bibfnamefont {K.~M.}\ \bibnamefont {Rudinger}}, \bibinfo {author} {\bibfnamefont {E.}~\bibnamefont {Nielsen}}, \bibinfo {author} {\bibfnamefont {K.~C.}\ \bibnamefont {Young}}, \bibinfo {author} {\bibfnamefont {T.~J.}\ \bibnamefont {Proctor}}, \bibinfo {author} {\bibfnamefont {A.~D.}\ \bibnamefont {Baczewski}}, \bibinfo {author} {\bibfnamefont {A.}~\bibnamefont {Laucht}}, \emph {et~al.},\ }\bibfield  {title} {\bibinfo {title} {Precision tomography of a three-qubit donor quantum processor in silicon},\ }\href@noop {} {\bibfield  {journal} {\bibinfo  {journal} {Nature}\ }\textbf {\bibinfo {volume} {601}},\ \bibinfo {pages} {348} (\bibinfo {year} {2022})}\BibitemShut {NoStop}%
\bibitem [{\citenamefont {Thorvaldson}\ \emph {et~al.}(2025)\citenamefont {Thorvaldson}, \citenamefont {Poulos}, \citenamefont {Moehle}, \citenamefont {Misha}, \citenamefont {Edlbauer}, \citenamefont {Reiner}, \citenamefont {Geng}, \citenamefont {Voisin}, \citenamefont {Jones}, \citenamefont {Donnelly} \emph {et~al.}}]{thorvaldson2025grover}%
  \BibitemOpen
  \bibfield  {author} {\bibinfo {author} {\bibfnamefont {I.}~\bibnamefont {Thorvaldson}}, \bibinfo {author} {\bibfnamefont {D.}~\bibnamefont {Poulos}}, \bibinfo {author} {\bibfnamefont {C.~M.}\ \bibnamefont {Moehle}}, \bibinfo {author} {\bibfnamefont {S.~H.}\ \bibnamefont {Misha}}, \bibinfo {author} {\bibfnamefont {H.}~\bibnamefont {Edlbauer}}, \bibinfo {author} {\bibfnamefont {J.}~\bibnamefont {Reiner}}, \bibinfo {author} {\bibfnamefont {H.}~\bibnamefont {Geng}}, \bibinfo {author} {\bibfnamefont {B.}~\bibnamefont {Voisin}}, \bibinfo {author} {\bibfnamefont {M.~T.}\ \bibnamefont {Jones}}, \bibinfo {author} {\bibfnamefont {M.~B.}\ \bibnamefont {Donnelly}}, \emph {et~al.},\ }\bibfield  {title} {\bibinfo {title} {Grover’s algorithm in a four-qubit silicon processor above the fault-tolerant threshold},\ }\href@noop {} {\bibfield  {journal} {\bibinfo  {journal} {Nature Nanotechnology}\ ,\ \bibinfo {pages} {1}} (\bibinfo {year} {2025})}\BibitemShut {NoStop}%
\bibitem [{\citenamefont {Edlbauer}\ \emph {et~al.}(2025)\citenamefont {Edlbauer}, \citenamefont {Wang}, \citenamefont {Huq}, \citenamefont {Thorvaldson}, \citenamefont {Jones}, \citenamefont {Misha}, \citenamefont {Pappas}, \citenamefont {Moehle}, \citenamefont {Hsueh}, \citenamefont {Bornemann} \emph {et~al.}}]{edlbauer202511}%
  \BibitemOpen
  \bibfield  {author} {\bibinfo {author} {\bibfnamefont {H.}~\bibnamefont {Edlbauer}}, \bibinfo {author} {\bibfnamefont {J.}~\bibnamefont {Wang}}, \bibinfo {author} {\bibfnamefont {A.~S.-E.}\ \bibnamefont {Huq}}, \bibinfo {author} {\bibfnamefont {I.}~\bibnamefont {Thorvaldson}}, \bibinfo {author} {\bibfnamefont {M.~T.}\ \bibnamefont {Jones}}, \bibinfo {author} {\bibfnamefont {S.~H.}\ \bibnamefont {Misha}}, \bibinfo {author} {\bibfnamefont {W.~J.}\ \bibnamefont {Pappas}}, \bibinfo {author} {\bibfnamefont {C.~M.}\ \bibnamefont {Moehle}}, \bibinfo {author} {\bibfnamefont {Y.-L.}\ \bibnamefont {Hsueh}}, \bibinfo {author} {\bibfnamefont {H.}~\bibnamefont {Bornemann}}, \emph {et~al.},\ }\bibfield  {title} {\bibinfo {title} {An 11-qubit atom processor in silicon},\ }\href@noop {} {\bibfield  {journal} {\bibinfo  {journal} {Nature}\ }\textbf {\bibinfo {volume} {648}},\ \bibinfo {pages} {569} (\bibinfo {year} {2025})}\BibitemShut {NoStop}%
\bibitem [{\citenamefont {Kalra}\ \emph {et~al.}(2014)\citenamefont {Kalra}, \citenamefont {Laucht}, \citenamefont {Hill},\ and\ \citenamefont {Morello}}]{kalra2014robust}%
  \BibitemOpen
  \bibfield  {author} {\bibinfo {author} {\bibfnamefont {R.}~\bibnamefont {Kalra}}, \bibinfo {author} {\bibfnamefont {A.}~\bibnamefont {Laucht}}, \bibinfo {author} {\bibfnamefont {C.~D.}\ \bibnamefont {Hill}},\ and\ \bibinfo {author} {\bibfnamefont {A.}~\bibnamefont {Morello}},\ }\bibfield  {title} {\bibinfo {title} {Robust two-qubit gates for donors in silicon controlled by hyperfine interactions},\ }\href@noop {} {\bibfield  {journal} {\bibinfo  {journal} {Physical Review X}\ }\textbf {\bibinfo {volume} {4}},\ \bibinfo {pages} {021044} (\bibinfo {year} {2014})}\BibitemShut {NoStop}%
\bibitem [{\citenamefont {M{\k{a}}dzik}\ \emph {et~al.}(2021)\citenamefont {M{\k{a}}dzik}, \citenamefont {Laucht}, \citenamefont {Hudson}, \citenamefont {Jakob}, \citenamefont {Johnson}, \citenamefont {Jamieson}, \citenamefont {Itoh}, \citenamefont {Dzurak},\ and\ \citenamefont {Morello}}]{madzik2021conditional}%
  \BibitemOpen
  \bibfield  {author} {\bibinfo {author} {\bibfnamefont {M.~T.}\ \bibnamefont {M{\k{a}}dzik}}, \bibinfo {author} {\bibfnamefont {A.}~\bibnamefont {Laucht}}, \bibinfo {author} {\bibfnamefont {F.~E.}\ \bibnamefont {Hudson}}, \bibinfo {author} {\bibfnamefont {A.~M.}\ \bibnamefont {Jakob}}, \bibinfo {author} {\bibfnamefont {B.~C.}\ \bibnamefont {Johnson}}, \bibinfo {author} {\bibfnamefont {D.~N.}\ \bibnamefont {Jamieson}}, \bibinfo {author} {\bibfnamefont {K.~M.}\ \bibnamefont {Itoh}}, \bibinfo {author} {\bibfnamefont {A.~S.}\ \bibnamefont {Dzurak}},\ and\ \bibinfo {author} {\bibfnamefont {A.}~\bibnamefont {Morello}},\ }\bibfield  {title} {\bibinfo {title} {Conditional quantum operation of two exchange-coupled single-donor spin qubits in a mos-compatible silicon device},\ }\href@noop {} {\bibfield  {journal} {\bibinfo  {journal} {Nature Communications}\ }\textbf {\bibinfo {volume} {12}},\ \bibinfo {pages} {181} (\bibinfo {year} {2021})}\BibitemShut {NoStop}%
\bibitem [{\citenamefont {Stemp}\ \emph {et~al.}(2025)\citenamefont {Stemp}, \citenamefont {van Blankenstein}, \citenamefont {Asaad}, \citenamefont {M{\k{a}}dzik}, \citenamefont {Joecker}, \citenamefont {Firgau}, \citenamefont {Laucht}, \citenamefont {Hudson}, \citenamefont {Dzurak}, \citenamefont {Itoh} \emph {et~al.}}]{stemp2025scalable}%
  \BibitemOpen
  \bibfield  {author} {\bibinfo {author} {\bibfnamefont {H.~G.}\ \bibnamefont {Stemp}}, \bibinfo {author} {\bibfnamefont {M.~R.}\ \bibnamefont {van Blankenstein}}, \bibinfo {author} {\bibfnamefont {S.}~\bibnamefont {Asaad}}, \bibinfo {author} {\bibfnamefont {M.~T.}\ \bibnamefont {M{\k{a}}dzik}}, \bibinfo {author} {\bibfnamefont {B.}~\bibnamefont {Joecker}}, \bibinfo {author} {\bibfnamefont {H.~R.}\ \bibnamefont {Firgau}}, \bibinfo {author} {\bibfnamefont {A.}~\bibnamefont {Laucht}}, \bibinfo {author} {\bibfnamefont {F.~E.}\ \bibnamefont {Hudson}}, \bibinfo {author} {\bibfnamefont {A.~S.}\ \bibnamefont {Dzurak}}, \bibinfo {author} {\bibfnamefont {K.~M.}\ \bibnamefont {Itoh}}, \emph {et~al.},\ }\bibfield  {title} {\bibinfo {title} {Scalable entanglement of nuclear spins mediated by electron exchange},\ }\href@noop {} {\bibfield  {journal} {\bibinfo  {journal} {Science}\ }\textbf {\bibinfo {volume} {389}},\ \bibinfo {pages} {1234} (\bibinfo {year} {2025})}\BibitemShut {NoStop}%
\bibitem [{\citenamefont {Morello}\ \emph {et~al.}(2010)\citenamefont {Morello}, \citenamefont {Pla}, \citenamefont {Zwanenburg}, \citenamefont {Chan}, \citenamefont {Tan}, \citenamefont {Huebl}, \citenamefont {M{\"o}tt{\"o}nen}, \citenamefont {Nugroho}, \citenamefont {Yang}, \citenamefont {Van~Donkelaar} \emph {et~al.}}]{morello2010single}%
  \BibitemOpen
  \bibfield  {author} {\bibinfo {author} {\bibfnamefont {A.}~\bibnamefont {Morello}}, \bibinfo {author} {\bibfnamefont {J.~J.}\ \bibnamefont {Pla}}, \bibinfo {author} {\bibfnamefont {F.~A.}\ \bibnamefont {Zwanenburg}}, \bibinfo {author} {\bibfnamefont {K.~W.}\ \bibnamefont {Chan}}, \bibinfo {author} {\bibfnamefont {K.~Y.}\ \bibnamefont {Tan}}, \bibinfo {author} {\bibfnamefont {H.}~\bibnamefont {Huebl}}, \bibinfo {author} {\bibfnamefont {M.}~\bibnamefont {M{\"o}tt{\"o}nen}}, \bibinfo {author} {\bibfnamefont {C.~D.}\ \bibnamefont {Nugroho}}, \bibinfo {author} {\bibfnamefont {C.}~\bibnamefont {Yang}}, \bibinfo {author} {\bibfnamefont {J.~A.}\ \bibnamefont {Van~Donkelaar}}, \emph {et~al.},\ }\bibfield  {title} {\bibinfo {title} {Single-shot readout of an electron spin in silicon},\ }\href@noop {} {\bibfield  {journal} {\bibinfo  {journal} {Nature}\ }\textbf {\bibinfo {volume} {467}},\ \bibinfo {pages} {687} (\bibinfo {year} {2010})}\BibitemShut {NoStop}%
\bibitem [{\citenamefont {Elzerman}\ \emph {et~al.}(2004)\citenamefont {Elzerman}, \citenamefont {Hanson}, \citenamefont {Willems~van Beveren}, \citenamefont {Witkamp}, \citenamefont {Vandersypen},\ and\ \citenamefont {Kouwenhoven}}]{elzerman2004single}%
  \BibitemOpen
  \bibfield  {author} {\bibinfo {author} {\bibfnamefont {J.}~\bibnamefont {Elzerman}}, \bibinfo {author} {\bibfnamefont {R.}~\bibnamefont {Hanson}}, \bibinfo {author} {\bibfnamefont {L.}~\bibnamefont {Willems~van Beveren}}, \bibinfo {author} {\bibfnamefont {B.}~\bibnamefont {Witkamp}}, \bibinfo {author} {\bibfnamefont {L.}~\bibnamefont {Vandersypen}},\ and\ \bibinfo {author} {\bibfnamefont {L.~P.}\ \bibnamefont {Kouwenhoven}},\ }\bibfield  {title} {\bibinfo {title} {Single-shot read-out of an individual electron spin in a quantum dot},\ }\href@noop {} {\bibfield  {journal} {\bibinfo  {journal} {Nature}\ }\textbf {\bibinfo {volume} {430}},\ \bibinfo {pages} {431} (\bibinfo {year} {2004})}\BibitemShut {NoStop}%
\bibitem [{\citenamefont {Jakob}\ \emph {et~al.}(2022)\citenamefont {Jakob}, \citenamefont {Robson}, \citenamefont {Schmitt}, \citenamefont {Mourik}, \citenamefont {Posselt}, \citenamefont {Spemann}, \citenamefont {Johnson}, \citenamefont {Firgau}, \citenamefont {Mayes}, \citenamefont {McCallum} \emph {et~al.}}]{jakob2022deterministic}%
  \BibitemOpen
  \bibfield  {author} {\bibinfo {author} {\bibfnamefont {A.~M.}\ \bibnamefont {Jakob}}, \bibinfo {author} {\bibfnamefont {S.~G.}\ \bibnamefont {Robson}}, \bibinfo {author} {\bibfnamefont {V.}~\bibnamefont {Schmitt}}, \bibinfo {author} {\bibfnamefont {V.}~\bibnamefont {Mourik}}, \bibinfo {author} {\bibfnamefont {M.}~\bibnamefont {Posselt}}, \bibinfo {author} {\bibfnamefont {D.}~\bibnamefont {Spemann}}, \bibinfo {author} {\bibfnamefont {B.~C.}\ \bibnamefont {Johnson}}, \bibinfo {author} {\bibfnamefont {H.~R.}\ \bibnamefont {Firgau}}, \bibinfo {author} {\bibfnamefont {E.}~\bibnamefont {Mayes}}, \bibinfo {author} {\bibfnamefont {J.~C.}\ \bibnamefont {McCallum}}, \emph {et~al.},\ }\bibfield  {title} {\bibinfo {title} {Deterministic shallow dopant implantation in silicon with detection confidence upper-bound to 99.85\% by ion--solid interactions},\ }\href@noop {} {\bibfield  {journal} {\bibinfo  {journal} {Advanced Materials}\ }\textbf {\bibinfo {volume} {34}},\ \bibinfo {pages} {2103235} (\bibinfo {year}
  {2022})}\BibitemShut {NoStop}%
\bibitem [{\citenamefont {Jamieson}(2013)}]{jamieson2013single}%
  \BibitemOpen
  \bibfield  {author} {\bibinfo {author} {\bibfnamefont {D.~N.}\ \bibnamefont {Jamieson}},\ }\bibfield  {title} {\bibinfo {title} {Single-ion implantation for quantum computing},\ }in\ \href@noop {} {\emph {\bibinfo {booktitle} {Single-Atom Nanoelectronics}}},\ \bibinfo {editor} {edited by\ \bibinfo {editor} {\bibfnamefont {E.}~\bibnamefont {Prati}}\ and\ \bibinfo {editor} {\bibfnamefont {T.}~\bibnamefont {Shinada}}}\ (\bibinfo  {publisher} {Routledge},\ \bibinfo {year} {2013})\BibitemShut {NoStop}%
\bibitem [{\citenamefont {Fuechsle}\ \emph {et~al.}(2012)\citenamefont {Fuechsle}, \citenamefont {Miwa}, \citenamefont {Mahapatra}, \citenamefont {Ryu}, \citenamefont {Lee}, \citenamefont {Warschkow}, \citenamefont {Hollenberg}, \citenamefont {Klimeck},\ and\ \citenamefont {Simmons}}]{fuechsle2012single}%
  \BibitemOpen
  \bibfield  {author} {\bibinfo {author} {\bibfnamefont {M.}~\bibnamefont {Fuechsle}}, \bibinfo {author} {\bibfnamefont {J.~A.}\ \bibnamefont {Miwa}}, \bibinfo {author} {\bibfnamefont {S.}~\bibnamefont {Mahapatra}}, \bibinfo {author} {\bibfnamefont {H.}~\bibnamefont {Ryu}}, \bibinfo {author} {\bibfnamefont {S.}~\bibnamefont {Lee}}, \bibinfo {author} {\bibfnamefont {O.}~\bibnamefont {Warschkow}}, \bibinfo {author} {\bibfnamefont {L.~C.}\ \bibnamefont {Hollenberg}}, \bibinfo {author} {\bibfnamefont {G.}~\bibnamefont {Klimeck}},\ and\ \bibinfo {author} {\bibfnamefont {M.~Y.}\ \bibnamefont {Simmons}},\ }\bibfield  {title} {\bibinfo {title} {A single-atom transistor},\ }\href@noop {} {\bibfield  {journal} {\bibinfo  {journal} {Nature Nanotechnology}\ }\textbf {\bibinfo {volume} {7}},\ \bibinfo {pages} {242} (\bibinfo {year} {2012})}\BibitemShut {NoStop}%
\bibitem [{\citenamefont {Schofield}\ \emph {et~al.}(2025)\citenamefont {Schofield}, \citenamefont {Fisher}, \citenamefont {Ginossar}, \citenamefont {Lyding}, \citenamefont {Silver}, \citenamefont {Fei}, \citenamefont {Namboodiri}, \citenamefont {Wyrick}, \citenamefont {Masteghin}, \citenamefont {Cox} \emph {et~al.}}]{schofield2025roadmap}%
  \BibitemOpen
  \bibfield  {author} {\bibinfo {author} {\bibfnamefont {S.~R.}\ \bibnamefont {Schofield}}, \bibinfo {author} {\bibfnamefont {A.~J.}\ \bibnamefont {Fisher}}, \bibinfo {author} {\bibfnamefont {E.}~\bibnamefont {Ginossar}}, \bibinfo {author} {\bibfnamefont {J.~W.}\ \bibnamefont {Lyding}}, \bibinfo {author} {\bibfnamefont {R.}~\bibnamefont {Silver}}, \bibinfo {author} {\bibfnamefont {F.}~\bibnamefont {Fei}}, \bibinfo {author} {\bibfnamefont {P.}~\bibnamefont {Namboodiri}}, \bibinfo {author} {\bibfnamefont {J.}~\bibnamefont {Wyrick}}, \bibinfo {author} {\bibfnamefont {M.~G.}\ \bibnamefont {Masteghin}}, \bibinfo {author} {\bibfnamefont {D.~C.}\ \bibnamefont {Cox}}, \emph {et~al.},\ }\bibfield  {title} {\bibinfo {title} {Roadmap on atomic-scale semiconductor devices},\ }\href@noop {} {\bibfield  {journal} {\bibinfo  {journal} {Nano Futures}\ }\textbf {\bibinfo {volume} {9}},\ \bibinfo {pages} {012001} (\bibinfo {year} {2025})}\BibitemShut {NoStop}%
\bibitem [{\citenamefont {Li}\ \emph {et~al.}(2010)\citenamefont {Li}, \citenamefont {Cywi{\'n}ski}, \citenamefont {Culcer}, \citenamefont {Hu},\ and\ \citenamefont {Das~Sarma}}]{li2010exchange}%
  \BibitemOpen
  \bibfield  {author} {\bibinfo {author} {\bibfnamefont {Q.}~\bibnamefont {Li}}, \bibinfo {author} {\bibfnamefont {{\L}.}~\bibnamefont {Cywi{\'n}ski}}, \bibinfo {author} {\bibfnamefont {D.}~\bibnamefont {Culcer}}, \bibinfo {author} {\bibfnamefont {X.}~\bibnamefont {Hu}},\ and\ \bibinfo {author} {\bibfnamefont {S.}~\bibnamefont {Das~Sarma}},\ }\bibfield  {title} {\bibinfo {title} {Exchange coupling in silicon quantum dots: Theoretical considerations for quantum computation},\ }\href@noop {} {\bibfield  {journal} {\bibinfo  {journal} {Physical Review B—Condensed Matter and Materials Physics}\ }\textbf {\bibinfo {volume} {81}},\ \bibinfo {pages} {085313} (\bibinfo {year} {2010})}\BibitemShut {NoStop}%
\bibitem [{\citenamefont {Burkard}\ \emph {et~al.}(1999)\citenamefont {Burkard}, \citenamefont {Loss},\ and\ \citenamefont {DiVincenzo}}]{burkard1999coupled}%
  \BibitemOpen
  \bibfield  {author} {\bibinfo {author} {\bibfnamefont {G.}~\bibnamefont {Burkard}}, \bibinfo {author} {\bibfnamefont {D.}~\bibnamefont {Loss}},\ and\ \bibinfo {author} {\bibfnamefont {D.~P.}\ \bibnamefont {DiVincenzo}},\ }\bibfield  {title} {\bibinfo {title} {Coupled quantum dots as quantum gates},\ }\href@noop {} {\bibfield  {journal} {\bibinfo  {journal} {Physical Review B}\ }\textbf {\bibinfo {volume} {59}},\ \bibinfo {pages} {2070} (\bibinfo {year} {1999})}\BibitemShut {NoStop}%
\bibitem [{\citenamefont {Koiller}\ \emph {et~al.}(2001)\citenamefont {Koiller}, \citenamefont {Hu},\ and\ \citenamefont {Sarma}}]{koiller2001exchange}%
  \BibitemOpen
  \bibfield  {author} {\bibinfo {author} {\bibfnamefont {B.}~\bibnamefont {Koiller}}, \bibinfo {author} {\bibfnamefont {X.}~\bibnamefont {Hu}},\ and\ \bibinfo {author} {\bibfnamefont {S.~D.}\ \bibnamefont {Sarma}},\ }\bibfield  {title} {\bibinfo {title} {Exchange in silicon-based quantum computer architecture},\ }\href@noop {} {\bibfield  {journal} {\bibinfo  {journal} {Physical Review Letters}\ }\textbf {\bibinfo {volume} {88}},\ \bibinfo {pages} {027903} (\bibinfo {year} {2001})}\BibitemShut {NoStop}%
\bibitem [{\citenamefont {Klymenko}\ and\ \citenamefont {Remacle}(2014)}]{klymenko2014electronic}%
  \BibitemOpen
  \bibfield  {author} {\bibinfo {author} {\bibfnamefont {M.}~\bibnamefont {Klymenko}}\ and\ \bibinfo {author} {\bibfnamefont {F.}~\bibnamefont {Remacle}},\ }\bibfield  {title} {\bibinfo {title} {Electronic states and wavefunctions of diatomic donor molecular ions in silicon: multi-valley envelope function theory},\ }\href@noop {} {\bibfield  {journal} {\bibinfo  {journal} {Journal of Physics: Condensed Matter}\ }\textbf {\bibinfo {volume} {26}},\ \bibinfo {pages} {065302} (\bibinfo {year} {2014})}\BibitemShut {NoStop}%
\bibitem [{\citenamefont {Saraiva}\ \emph {et~al.}(2015)\citenamefont {Saraiva}, \citenamefont {Baena}, \citenamefont {Calder{\'o}n},\ and\ \citenamefont {Koiller}}]{saraiva2015theory}%
  \BibitemOpen
  \bibfield  {author} {\bibinfo {author} {\bibfnamefont {A.}~\bibnamefont {Saraiva}}, \bibinfo {author} {\bibfnamefont {A.}~\bibnamefont {Baena}}, \bibinfo {author} {\bibfnamefont {M.}~\bibnamefont {Calder{\'o}n}},\ and\ \bibinfo {author} {\bibfnamefont {B.}~\bibnamefont {Koiller}},\ }\bibfield  {title} {\bibinfo {title} {Theory of one and two donors in silicon},\ }\href@noop {} {\bibfield  {journal} {\bibinfo  {journal} {Journal of Physics: Condensed Matter}\ }\textbf {\bibinfo {volume} {27}},\ \bibinfo {pages} {154208} (\bibinfo {year} {2015})}\BibitemShut {NoStop}%
\bibitem [{\citenamefont {Joecker}\ \emph {et~al.}(2021)\citenamefont {Joecker}, \citenamefont {Baczewski}, \citenamefont {Gamble}, \citenamefont {Pla}, \citenamefont {Saraiva},\ and\ \citenamefont {Morello}}]{joecker2021full}%
  \BibitemOpen
  \bibfield  {author} {\bibinfo {author} {\bibfnamefont {B.}~\bibnamefont {Joecker}}, \bibinfo {author} {\bibfnamefont {A.~D.}\ \bibnamefont {Baczewski}}, \bibinfo {author} {\bibfnamefont {J.~K.}\ \bibnamefont {Gamble}}, \bibinfo {author} {\bibfnamefont {J.~J.}\ \bibnamefont {Pla}}, \bibinfo {author} {\bibfnamefont {A.}~\bibnamefont {Saraiva}},\ and\ \bibinfo {author} {\bibfnamefont {A.}~\bibnamefont {Morello}},\ }\bibfield  {title} {\bibinfo {title} {Full configuration interaction simulations of exchange-coupled donors in silicon using multi-valley effective mass theory},\ }\href@noop {} {\bibfield  {journal} {\bibinfo  {journal} {New Journal of Physics}\ }\textbf {\bibinfo {volume} {23}},\ \bibinfo {pages} {073007} (\bibinfo {year} {2021})}\BibitemShut {NoStop}%
\bibitem [{\citenamefont {Pla}\ \emph {et~al.}(2012)\citenamefont {Pla}, \citenamefont {Tan}, \citenamefont {Dehollain}, \citenamefont {Lim}, \citenamefont {Morton}, \citenamefont {Jamieson}, \citenamefont {Dzurak},\ and\ \citenamefont {Morello}}]{pla2012single}%
  \BibitemOpen
  \bibfield  {author} {\bibinfo {author} {\bibfnamefont {J.~J.}\ \bibnamefont {Pla}}, \bibinfo {author} {\bibfnamefont {K.~Y.}\ \bibnamefont {Tan}}, \bibinfo {author} {\bibfnamefont {J.~P.}\ \bibnamefont {Dehollain}}, \bibinfo {author} {\bibfnamefont {W.~H.}\ \bibnamefont {Lim}}, \bibinfo {author} {\bibfnamefont {J.~J.}\ \bibnamefont {Morton}}, \bibinfo {author} {\bibfnamefont {D.~N.}\ \bibnamefont {Jamieson}}, \bibinfo {author} {\bibfnamefont {A.~S.}\ \bibnamefont {Dzurak}},\ and\ \bibinfo {author} {\bibfnamefont {A.}~\bibnamefont {Morello}},\ }\bibfield  {title} {\bibinfo {title} {A single-atom electron spin qubit in silicon},\ }\href@noop {} {\bibfield  {journal} {\bibinfo  {journal} {Nature}\ }\textbf {\bibinfo {volume} {489}},\ \bibinfo {pages} {541} (\bibinfo {year} {2012})}\BibitemShut {NoStop}%
\bibitem [{\citenamefont {Dehollain}\ \emph {et~al.}(2016{\natexlab{b}})\citenamefont {Dehollain}, \citenamefont {Simmons}, \citenamefont {Muhonen}, \citenamefont {Kalra}, \citenamefont {Laucht}, \citenamefont {Hudson}, \citenamefont {Itoh}, \citenamefont {Jamieson}, \citenamefont {McCallum}, \citenamefont {Dzurak} \emph {et~al.}}]{dehollain2016bell}%
  \BibitemOpen
  \bibfield  {author} {\bibinfo {author} {\bibfnamefont {J.~P.}\ \bibnamefont {Dehollain}}, \bibinfo {author} {\bibfnamefont {S.}~\bibnamefont {Simmons}}, \bibinfo {author} {\bibfnamefont {J.~T.}\ \bibnamefont {Muhonen}}, \bibinfo {author} {\bibfnamefont {R.}~\bibnamefont {Kalra}}, \bibinfo {author} {\bibfnamefont {A.}~\bibnamefont {Laucht}}, \bibinfo {author} {\bibfnamefont {F.}~\bibnamefont {Hudson}}, \bibinfo {author} {\bibfnamefont {K.~M.}\ \bibnamefont {Itoh}}, \bibinfo {author} {\bibfnamefont {D.~N.}\ \bibnamefont {Jamieson}}, \bibinfo {author} {\bibfnamefont {J.~C.}\ \bibnamefont {McCallum}}, \bibinfo {author} {\bibfnamefont {A.~S.}\ \bibnamefont {Dzurak}}, \emph {et~al.},\ }\bibfield  {title} {\bibinfo {title} {Bell's inequality violation with spins in silicon},\ }\href@noop {} {\bibfield  {journal} {\bibinfo  {journal} {Nature Nanotechnology}\ }\textbf {\bibinfo {volume} {11}},\ \bibinfo {pages} {242} (\bibinfo {year} {2016}{\natexlab{b}})}\BibitemShut {NoStop}%
\bibitem [{\citenamefont {Huang}\ \emph {et~al.}(2019)\citenamefont {Huang}, \citenamefont {Yang}, \citenamefont {Chan}, \citenamefont {Tanttu}, \citenamefont {Hensen}, \citenamefont {Leon}, \citenamefont {Fogarty}, \citenamefont {Hwang}, \citenamefont {Hudson}, \citenamefont {Itoh} \emph {et~al.}}]{huang2019fidelity}%
  \BibitemOpen
  \bibfield  {author} {\bibinfo {author} {\bibfnamefont {W.}~\bibnamefont {Huang}}, \bibinfo {author} {\bibfnamefont {C.}~\bibnamefont {Yang}}, \bibinfo {author} {\bibfnamefont {K.}~\bibnamefont {Chan}}, \bibinfo {author} {\bibfnamefont {T.}~\bibnamefont {Tanttu}}, \bibinfo {author} {\bibfnamefont {B.}~\bibnamefont {Hensen}}, \bibinfo {author} {\bibfnamefont {R.}~\bibnamefont {Leon}}, \bibinfo {author} {\bibfnamefont {M.}~\bibnamefont {Fogarty}}, \bibinfo {author} {\bibfnamefont {J.}~\bibnamefont {Hwang}}, \bibinfo {author} {\bibfnamefont {F.}~\bibnamefont {Hudson}}, \bibinfo {author} {\bibfnamefont {K.~M.}\ \bibnamefont {Itoh}}, \emph {et~al.},\ }\bibfield  {title} {\bibinfo {title} {Fidelity benchmarks for two-qubit gates in silicon},\ }\href@noop {} {\bibfield  {journal} {\bibinfo  {journal} {Nature}\ }\textbf {\bibinfo {volume} {569}},\ \bibinfo {pages} {532} (\bibinfo {year} {2019})}\BibitemShut {NoStop}%
\bibitem [{\citenamefont {Feher}(1959)}]{feher1959electron}%
  \BibitemOpen
  \bibfield  {author} {\bibinfo {author} {\bibfnamefont {G.}~\bibnamefont {Feher}},\ }\bibfield  {title} {\bibinfo {title} {Electron spin resonance experiments on donors in silicon. i. electronic structure of donors by the electron nuclear double resonance technique},\ }\href@noop {} {\bibfield  {journal} {\bibinfo  {journal} {Physical Review}\ }\textbf {\bibinfo {volume} {114}},\ \bibinfo {pages} {1219} (\bibinfo {year} {1959})}\BibitemShut {NoStop}%
\bibitem [{\citenamefont {Usman}\ \emph {et~al.}(2015)\citenamefont {Usman}, \citenamefont {Hill}, \citenamefont {Rahman}, \citenamefont {Klimeck}, \citenamefont {Simmons}, \citenamefont {Rogge},\ and\ \citenamefont {Hollenberg}}]{usman2015strain}%
  \BibitemOpen
  \bibfield  {author} {\bibinfo {author} {\bibfnamefont {M.}~\bibnamefont {Usman}}, \bibinfo {author} {\bibfnamefont {C.~D.}\ \bibnamefont {Hill}}, \bibinfo {author} {\bibfnamefont {R.}~\bibnamefont {Rahman}}, \bibinfo {author} {\bibfnamefont {G.}~\bibnamefont {Klimeck}}, \bibinfo {author} {\bibfnamefont {M.~Y.}\ \bibnamefont {Simmons}}, \bibinfo {author} {\bibfnamefont {S.}~\bibnamefont {Rogge}},\ and\ \bibinfo {author} {\bibfnamefont {L.~C.}\ \bibnamefont {Hollenberg}},\ }\bibfield  {title} {\bibinfo {title} {Strain and electric field control of hyperfine interactions for donor spin qubits in silicon},\ }\href@noop {} {\bibfield  {journal} {\bibinfo  {journal} {Physical Review B}\ }\textbf {\bibinfo {volume} {91}},\ \bibinfo {pages} {245209} (\bibinfo {year} {2015})}\BibitemShut {NoStop}%
\bibitem [{\citenamefont {Mansir}\ \emph {et~al.}(2018)\citenamefont {Mansir}, \citenamefont {Conti}, \citenamefont {Zeng}, \citenamefont {Pla}, \citenamefont {Bertet}, \citenamefont {Swift}, \citenamefont {Van~de Walle}, \citenamefont {Thewalt}, \citenamefont {Sklenard}, \citenamefont {Niquet} \emph {et~al.}}]{mansir2018linear}%
  \BibitemOpen
  \bibfield  {author} {\bibinfo {author} {\bibfnamefont {J.}~\bibnamefont {Mansir}}, \bibinfo {author} {\bibfnamefont {P.}~\bibnamefont {Conti}}, \bibinfo {author} {\bibfnamefont {Z.}~\bibnamefont {Zeng}}, \bibinfo {author} {\bibfnamefont {J.~J.}\ \bibnamefont {Pla}}, \bibinfo {author} {\bibfnamefont {P.}~\bibnamefont {Bertet}}, \bibinfo {author} {\bibfnamefont {M.~W.}\ \bibnamefont {Swift}}, \bibinfo {author} {\bibfnamefont {C.~G.}\ \bibnamefont {Van~de Walle}}, \bibinfo {author} {\bibfnamefont {M.~L.}\ \bibnamefont {Thewalt}}, \bibinfo {author} {\bibfnamefont {B.}~\bibnamefont {Sklenard}}, \bibinfo {author} {\bibfnamefont {Y.-M.}\ \bibnamefont {Niquet}}, \emph {et~al.},\ }\bibfield  {title} {\bibinfo {title} {Linear hyperfine tuning of donor spins in silicon using hydrostatic strain},\ }\href@noop {} {\bibfield  {journal} {\bibinfo  {journal} {Physical Review Letters}\ }\textbf {\bibinfo {volume} {120}},\ \bibinfo {pages} {167701} (\bibinfo {year} {2018})}\BibitemShut {NoStop}%
\bibitem [{\citenamefont {Pla}\ \emph {et~al.}(2018)\citenamefont {Pla}, \citenamefont {Bienfait}, \citenamefont {Pica}, \citenamefont {Mansir}, \citenamefont {Mohiyaddin}, \citenamefont {Zeng}, \citenamefont {Morello}, \citenamefont {Schenkel}, \citenamefont {Morton} \emph {et~al.}}]{pla2018strain}%
  \BibitemOpen
  \bibfield  {author} {\bibinfo {author} {\bibfnamefont {J.}~\bibnamefont {Pla}}, \bibinfo {author} {\bibfnamefont {A.}~\bibnamefont {Bienfait}}, \bibinfo {author} {\bibfnamefont {G.}~\bibnamefont {Pica}}, \bibinfo {author} {\bibfnamefont {J.}~\bibnamefont {Mansir}}, \bibinfo {author} {\bibfnamefont {F.}~\bibnamefont {Mohiyaddin}}, \bibinfo {author} {\bibfnamefont {Z.}~\bibnamefont {Zeng}}, \bibinfo {author} {\bibfnamefont {A.}~\bibnamefont {Morello}}, \bibinfo {author} {\bibfnamefont {T.}~\bibnamefont {Schenkel}}, \bibinfo {author} {\bibfnamefont {J.}~\bibnamefont {Morton}}, \emph {et~al.},\ }\bibfield  {title} {\bibinfo {title} {Strain-induced spin-resonance shifts in silicon devices},\ }\href@noop {} {\bibfield  {journal} {\bibinfo  {journal} {Physical Review Applied}\ }\textbf {\bibinfo {volume} {9}},\ \bibinfo {pages} {044014} (\bibinfo {year} {2018})}\BibitemShut {NoStop}%
\bibitem [{\citenamefont {Rahman}\ \emph {et~al.}(2007)\citenamefont {Rahman}, \citenamefont {Wellard}, \citenamefont {Bradbury}, \citenamefont {Prada}, \citenamefont {Cole}, \citenamefont {Klimeck},\ and\ \citenamefont {Hollenberg}}]{rahman2007high}%
  \BibitemOpen
  \bibfield  {author} {\bibinfo {author} {\bibfnamefont {R.}~\bibnamefont {Rahman}}, \bibinfo {author} {\bibfnamefont {C.~J.}\ \bibnamefont {Wellard}}, \bibinfo {author} {\bibfnamefont {F.~R.}\ \bibnamefont {Bradbury}}, \bibinfo {author} {\bibfnamefont {M.}~\bibnamefont {Prada}}, \bibinfo {author} {\bibfnamefont {J.~H.}\ \bibnamefont {Cole}}, \bibinfo {author} {\bibfnamefont {G.}~\bibnamefont {Klimeck}},\ and\ \bibinfo {author} {\bibfnamefont {L.~C.}\ \bibnamefont {Hollenberg}},\ }\bibfield  {title} {\bibinfo {title} {High precision quantum control of single donor spins in silicon},\ }\href@noop {} {\bibfield  {journal} {\bibinfo  {journal} {Physical Review Letters}\ }\textbf {\bibinfo {volume} {99}},\ \bibinfo {pages} {036403} (\bibinfo {year} {2007})}\BibitemShut {NoStop}%
\bibitem [{\citenamefont {Pica}\ \emph {et~al.}(2014)\citenamefont {Pica}, \citenamefont {Wolfowicz}, \citenamefont {Urdampilleta}, \citenamefont {Thewalt}, \citenamefont {Riemann}, \citenamefont {Abrosimov}, \citenamefont {Becker}, \citenamefont {Pohl}, \citenamefont {Morton}, \citenamefont {Bhatt} \emph {et~al.}}]{pica2014hyperfine}%
  \BibitemOpen
  \bibfield  {author} {\bibinfo {author} {\bibfnamefont {G.}~\bibnamefont {Pica}}, \bibinfo {author} {\bibfnamefont {G.}~\bibnamefont {Wolfowicz}}, \bibinfo {author} {\bibfnamefont {M.}~\bibnamefont {Urdampilleta}}, \bibinfo {author} {\bibfnamefont {M.~L.}\ \bibnamefont {Thewalt}}, \bibinfo {author} {\bibfnamefont {H.}~\bibnamefont {Riemann}}, \bibinfo {author} {\bibfnamefont {N.~V.}\ \bibnamefont {Abrosimov}}, \bibinfo {author} {\bibfnamefont {P.}~\bibnamefont {Becker}}, \bibinfo {author} {\bibfnamefont {H.-J.}\ \bibnamefont {Pohl}}, \bibinfo {author} {\bibfnamefont {J.~J.}\ \bibnamefont {Morton}}, \bibinfo {author} {\bibfnamefont {R.~N.}\ \bibnamefont {Bhatt}}, \emph {et~al.},\ }\bibfield  {title} {\bibinfo {title} {Hyperfine stark effect of shallow donors in silicon},\ }\href@noop {} {\bibfield  {journal} {\bibinfo  {journal} {Physical Review B}\ }\textbf {\bibinfo {volume} {90}},\ \bibinfo {pages} {195204} (\bibinfo {year} {2014})}\BibitemShut {NoStop}%
\bibitem [{\citenamefont {Laucht}\ \emph {et~al.}(2015)\citenamefont {Laucht}, \citenamefont {Muhonen}, \citenamefont {Mohiyaddin}, \citenamefont {Kalra}, \citenamefont {Dehollain}, \citenamefont {Freer}, \citenamefont {Hudson}, \citenamefont {Veldhorst}, \citenamefont {Rahman}, \citenamefont {Klimeck} \emph {et~al.}}]{laucht2015electrically}%
  \BibitemOpen
  \bibfield  {author} {\bibinfo {author} {\bibfnamefont {A.}~\bibnamefont {Laucht}}, \bibinfo {author} {\bibfnamefont {J.~T.}\ \bibnamefont {Muhonen}}, \bibinfo {author} {\bibfnamefont {F.~A.}\ \bibnamefont {Mohiyaddin}}, \bibinfo {author} {\bibfnamefont {R.}~\bibnamefont {Kalra}}, \bibinfo {author} {\bibfnamefont {J.~P.}\ \bibnamefont {Dehollain}}, \bibinfo {author} {\bibfnamefont {S.}~\bibnamefont {Freer}}, \bibinfo {author} {\bibfnamefont {F.~E.}\ \bibnamefont {Hudson}}, \bibinfo {author} {\bibfnamefont {M.}~\bibnamefont {Veldhorst}}, \bibinfo {author} {\bibfnamefont {R.}~\bibnamefont {Rahman}}, \bibinfo {author} {\bibfnamefont {G.}~\bibnamefont {Klimeck}}, \emph {et~al.},\ }\bibfield  {title} {\bibinfo {title} {Electrically controlling single-spin qubits in a continuous microwave field},\ }\href@noop {} {\bibfield  {journal} {\bibinfo  {journal} {Science Advances}\ }\textbf {\bibinfo {volume} {1}},\ \bibinfo {pages} {e1500022} (\bibinfo {year} {2015})}\BibitemShut {NoStop}%
\bibitem [{\citenamefont {Watson}\ \emph {et~al.}(2015)\citenamefont {Watson}, \citenamefont {Weber}, \citenamefont {House}, \citenamefont {B{\"u}ch},\ and\ \citenamefont {Simmons}}]{watson2015high}%
  \BibitemOpen
  \bibfield  {author} {\bibinfo {author} {\bibfnamefont {T.~F.}\ \bibnamefont {Watson}}, \bibinfo {author} {\bibfnamefont {B.}~\bibnamefont {Weber}}, \bibinfo {author} {\bibfnamefont {M.~G.}\ \bibnamefont {House}}, \bibinfo {author} {\bibfnamefont {H.}~\bibnamefont {B{\"u}ch}},\ and\ \bibinfo {author} {\bibfnamefont {M.~Y.}\ \bibnamefont {Simmons}},\ }\bibfield  {title} {\bibinfo {title} {High-fidelity rapid initialization and read-out of an electron spin via the single donor d-charge state},\ }\href@noop {} {\bibfield  {journal} {\bibinfo  {journal} {Physical Review Letters}\ }\textbf {\bibinfo {volume} {115}},\ \bibinfo {pages} {166806} (\bibinfo {year} {2015})}\BibitemShut {NoStop}%
\bibitem [{\citenamefont {Hanson}\ \emph {et~al.}(2005)\citenamefont {Hanson}, \citenamefont {van Beveren}, \citenamefont {Vink}, \citenamefont {Elzerman}, \citenamefont {Naber}, \citenamefont {Koppens}, \citenamefont {Kouwenhoven},\ and\ \citenamefont {Vandersypen}}]{hanson2005single}%
  \BibitemOpen
  \bibfield  {author} {\bibinfo {author} {\bibfnamefont {R.}~\bibnamefont {Hanson}}, \bibinfo {author} {\bibfnamefont {L.~W.}\ \bibnamefont {van Beveren}}, \bibinfo {author} {\bibfnamefont {I.}~\bibnamefont {Vink}}, \bibinfo {author} {\bibfnamefont {J.}~\bibnamefont {Elzerman}}, \bibinfo {author} {\bibfnamefont {W.}~\bibnamefont {Naber}}, \bibinfo {author} {\bibfnamefont {F.}~\bibnamefont {Koppens}}, \bibinfo {author} {\bibfnamefont {.~f.~L.}\ \bibnamefont {Kouwenhoven}},\ and\ \bibinfo {author} {\bibfnamefont {L.}~\bibnamefont {Vandersypen}},\ }\bibfield  {title} {\bibinfo {title} {Single-shot readout of electron spin states in a quantum dot using spin-dependent tunnel rates},\ }\href@noop {} {\bibfield  {journal} {\bibinfo  {journal} {Physical Review Letters}\ }\textbf {\bibinfo {volume} {94}},\ \bibinfo {pages} {196802} (\bibinfo {year} {2005})}\BibitemShut {NoStop}%
\bibitem [{\citenamefont {Dehollain}\ \emph {et~al.}(2014)\citenamefont {Dehollain}, \citenamefont {Muhonen}, \citenamefont {Tan}, \citenamefont {Saraiva}, \citenamefont {Jamieson}, \citenamefont {Dzurak},\ and\ \citenamefont {Morello}}]{dehollain2014single}%
  \BibitemOpen
  \bibfield  {author} {\bibinfo {author} {\bibfnamefont {J.~P.}\ \bibnamefont {Dehollain}}, \bibinfo {author} {\bibfnamefont {J.~T.}\ \bibnamefont {Muhonen}}, \bibinfo {author} {\bibfnamefont {K.~Y.}\ \bibnamefont {Tan}}, \bibinfo {author} {\bibfnamefont {A.}~\bibnamefont {Saraiva}}, \bibinfo {author} {\bibfnamefont {D.~N.}\ \bibnamefont {Jamieson}}, \bibinfo {author} {\bibfnamefont {A.~S.}\ \bibnamefont {Dzurak}},\ and\ \bibinfo {author} {\bibfnamefont {A.}~\bibnamefont {Morello}},\ }\bibfield  {title} {\bibinfo {title} {Single-shot readout and relaxation of singlet and triplet states in exchange-coupled p 31 electron spins in silicon},\ }\href@noop {} {\bibfield  {journal} {\bibinfo  {journal} {Physical Review Letters}\ }\textbf {\bibinfo {volume} {112}},\ \bibinfo {pages} {236801} (\bibinfo {year} {2014})}\BibitemShut {NoStop}%
\bibitem [{\citenamefont {Laucht}\ \emph {et~al.}(2014)\citenamefont {Laucht}, \citenamefont {Kalra}, \citenamefont {Muhonen}, \citenamefont {Dehollain}, \citenamefont {Mohiyaddin}, \citenamefont {Hudson}, \citenamefont {McCallum}, \citenamefont {Jamieson}, \citenamefont {Dzurak},\ and\ \citenamefont {Morello}}]{laucht2014high}%
  \BibitemOpen
  \bibfield  {author} {\bibinfo {author} {\bibfnamefont {A.}~\bibnamefont {Laucht}}, \bibinfo {author} {\bibfnamefont {R.}~\bibnamefont {Kalra}}, \bibinfo {author} {\bibfnamefont {J.~T.}\ \bibnamefont {Muhonen}}, \bibinfo {author} {\bibfnamefont {J.~P.}\ \bibnamefont {Dehollain}}, \bibinfo {author} {\bibfnamefont {F.~A.}\ \bibnamefont {Mohiyaddin}}, \bibinfo {author} {\bibfnamefont {F.}~\bibnamefont {Hudson}}, \bibinfo {author} {\bibfnamefont {J.~C.}\ \bibnamefont {McCallum}}, \bibinfo {author} {\bibfnamefont {D.~N.}\ \bibnamefont {Jamieson}}, \bibinfo {author} {\bibfnamefont {A.~S.}\ \bibnamefont {Dzurak}},\ and\ \bibinfo {author} {\bibfnamefont {A.}~\bibnamefont {Morello}},\ }\bibfield  {title} {\bibinfo {title} {High-fidelity adiabatic inversion of a $^{31}\text{P}$ electron spin qubit in natural silicon},\ }\href@noop {} {\bibfield  {journal} {\bibinfo  {journal} {Applied Physics Letters}\ }\textbf {\bibinfo {volume} {104}} (\bibinfo {year} {2014})}\BibitemShut {NoStop}%
\bibitem [{\citenamefont {Stemp}\ \emph {et~al.}(2026)\citenamefont {Stemp}, \citenamefont {van Blankenstein}, \citenamefont {Wilhelm}, \citenamefont {Asaad}, \citenamefont {Madzik}, \citenamefont {Laucht}, \citenamefont {Hudson}, \citenamefont {Dzurak}, \citenamefont {Itoh}, \citenamefont {Jakob}, \citenamefont {Johnson}, \citenamefont {Jamieson},\ and\ \citenamefont {Morello}}]{stemp2026paralleldata}%
  \BibitemOpen
  \bibfield  {author} {\bibinfo {author} {\bibfnamefont {H.}~\bibnamefont {Stemp}}, \bibinfo {author} {\bibfnamefont {M.}~\bibnamefont {van Blankenstein}}, \bibinfo {author} {\bibfnamefont {B.}~\bibnamefont {Wilhelm}}, \bibinfo {author} {\bibfnamefont {S.}~\bibnamefont {Asaad}}, \bibinfo {author} {\bibfnamefont {M.}~\bibnamefont {Madzik}}, \bibinfo {author} {\bibfnamefont {A.}~\bibnamefont {Laucht}}, \bibinfo {author} {\bibfnamefont {F.}~\bibnamefont {Hudson}}, \bibinfo {author} {\bibfnamefont {A.}~\bibnamefont {Dzurak}}, \bibinfo {author} {\bibfnamefont {K.}~\bibnamefont {Itoh}}, \bibinfo {author} {\bibfnamefont {A.}~\bibnamefont {Jakob}}, \bibinfo {author} {\bibfnamefont {B.}~\bibnamefont {Johnson}}, \bibinfo {author} {\bibfnamefont {D.}~\bibnamefont {Jamieson}},\ and\ \bibinfo {author} {\bibfnamefont {A.}~\bibnamefont {Morello}},\ }\bibfield  {title} {\bibinfo {title} {Dataset for electron readout contrast enhancement in the parallel nuclear regime of an exchange-coupled donor spin qubit system},\ }\href
  {https://doi.org/10.5061/dryad.1c59zw4b2} {https://doi.org/10.5061/dryad.1c59zw4b2} (\bibinfo {year} {2026})\BibitemShut {NoStop}%
\bibitem [{\citenamefont {Morello}\ \emph {et~al.}(2009)\citenamefont {Morello}, \citenamefont {Escott}, \citenamefont {Huebl}, \citenamefont {Willems~van Beveren}, \citenamefont {Hollenberg}, \citenamefont {Jamieson}, \citenamefont {Dzurak},\ and\ \citenamefont {Clark}}]{morello2009architecture}%
  \BibitemOpen
  \bibfield  {author} {\bibinfo {author} {\bibfnamefont {A.}~\bibnamefont {Morello}}, \bibinfo {author} {\bibfnamefont {C.}~\bibnamefont {Escott}}, \bibinfo {author} {\bibfnamefont {H.}~\bibnamefont {Huebl}}, \bibinfo {author} {\bibfnamefont {L.}~\bibnamefont {Willems~van Beveren}}, \bibinfo {author} {\bibfnamefont {L.}~\bibnamefont {Hollenberg}}, \bibinfo {author} {\bibfnamefont {D.}~\bibnamefont {Jamieson}}, \bibinfo {author} {\bibfnamefont {A.}~\bibnamefont {Dzurak}},\ and\ \bibinfo {author} {\bibfnamefont {R.}~\bibnamefont {Clark}},\ }\bibfield  {title} {\bibinfo {title} {Architecture for high-sensitivity single-shot readout and control of the electron spin of individual donors in silicon},\ }\href@noop {} {\bibfield  {journal} {\bibinfo  {journal} {Physical Review B}\ }\textbf {\bibinfo {volume} {80}},\ \bibinfo {pages} {081307} (\bibinfo {year} {2009})}\BibitemShut {NoStop}%
\bibitem [{\citenamefont {B{\"u}ch}\ \emph {et~al.}(2013)\citenamefont {B{\"u}ch}, \citenamefont {Mahapatra}, \citenamefont {Rahman}, \citenamefont {Morello},\ and\ \citenamefont {Simmons}}]{buch2013spin}%
  \BibitemOpen
  \bibfield  {author} {\bibinfo {author} {\bibfnamefont {H.}~\bibnamefont {B{\"u}ch}}, \bibinfo {author} {\bibfnamefont {S.}~\bibnamefont {Mahapatra}}, \bibinfo {author} {\bibfnamefont {R.}~\bibnamefont {Rahman}}, \bibinfo {author} {\bibfnamefont {A.}~\bibnamefont {Morello}},\ and\ \bibinfo {author} {\bibfnamefont {M.}~\bibnamefont {Simmons}},\ }\bibfield  {title} {\bibinfo {title} {Spin readout and addressability of phosphorus-donor clusters in silicon},\ }\href@noop {} {\bibfield  {journal} {\bibinfo  {journal} {Nature Communications}\ }\textbf {\bibinfo {volume} {4}},\ \bibinfo {pages} {2017} (\bibinfo {year} {2013})}\BibitemShut {NoStop}%
\bibitem [{\citenamefont {Mohiyaddin}\ \emph {et~al.}(2013)\citenamefont {Mohiyaddin}, \citenamefont {Rahman}, \citenamefont {Kalra}, \citenamefont {Klimeck}, \citenamefont {Hollenberg}, \citenamefont {Pla}, \citenamefont {Dzurak},\ and\ \citenamefont {Morello}}]{mohiyaddin2013noninvasive}%
  \BibitemOpen
  \bibfield  {author} {\bibinfo {author} {\bibfnamefont {F.~A.}\ \bibnamefont {Mohiyaddin}}, \bibinfo {author} {\bibfnamefont {R.}~\bibnamefont {Rahman}}, \bibinfo {author} {\bibfnamefont {R.}~\bibnamefont {Kalra}}, \bibinfo {author} {\bibfnamefont {G.}~\bibnamefont {Klimeck}}, \bibinfo {author} {\bibfnamefont {L.~C.}\ \bibnamefont {Hollenberg}}, \bibinfo {author} {\bibfnamefont {J.~J.}\ \bibnamefont {Pla}}, \bibinfo {author} {\bibfnamefont {A.~S.}\ \bibnamefont {Dzurak}},\ and\ \bibinfo {author} {\bibfnamefont {A.}~\bibnamefont {Morello}},\ }\bibfield  {title} {\bibinfo {title} {Noninvasive spatial metrology of single-atom devices},\ }\href@noop {} {\bibfield  {journal} {\bibinfo  {journal} {Nano Letters}\ }\textbf {\bibinfo {volume} {13}},\ \bibinfo {pages} {1903} (\bibinfo {year} {2013})}\BibitemShut {NoStop}%
\bibitem [{\citenamefont {Johnson}\ \emph {et~al.}(2022)\citenamefont {Johnson}, \citenamefont {M{\k{a}}dzik}, \citenamefont {Hudson}, \citenamefont {Itoh}, \citenamefont {Jakob}, \citenamefont {Jamieson}, \citenamefont {Dzurak},\ and\ \citenamefont {Morello}}]{johnson2022beating}%
  \BibitemOpen
  \bibfield  {author} {\bibinfo {author} {\bibfnamefont {M.~A.}\ \bibnamefont {Johnson}}, \bibinfo {author} {\bibfnamefont {M.~T.}\ \bibnamefont {M{\k{a}}dzik}}, \bibinfo {author} {\bibfnamefont {F.~E.}\ \bibnamefont {Hudson}}, \bibinfo {author} {\bibfnamefont {K.~M.}\ \bibnamefont {Itoh}}, \bibinfo {author} {\bibfnamefont {A.~M.}\ \bibnamefont {Jakob}}, \bibinfo {author} {\bibfnamefont {D.~N.}\ \bibnamefont {Jamieson}}, \bibinfo {author} {\bibfnamefont {A.}~\bibnamefont {Dzurak}},\ and\ \bibinfo {author} {\bibfnamefont {A.}~\bibnamefont {Morello}},\ }\bibfield  {title} {\bibinfo {title} {Beating the thermal limit of qubit initialization with a bayesian maxwell’s demon},\ }\href@noop {} {\bibfield  {journal} {\bibinfo  {journal} {Physical Review X}\ }\textbf {\bibinfo {volume} {12}},\ \bibinfo {pages} {041008} (\bibinfo {year} {2022})}\BibitemShut {NoStop}%
\bibitem [{\citenamefont {Huang}\ \emph {et~al.}(2021)\citenamefont {Huang}, \citenamefont {Lim}, \citenamefont {Leon}, \citenamefont {Yang}, \citenamefont {Hudson}, \citenamefont {Escott}, \citenamefont {Saraiva}, \citenamefont {Dzurak},\ and\ \citenamefont {Laucht}}]{huang2021high}%
  \BibitemOpen
  \bibfield  {author} {\bibinfo {author} {\bibfnamefont {J.~Y.}\ \bibnamefont {Huang}}, \bibinfo {author} {\bibfnamefont {W.~H.}\ \bibnamefont {Lim}}, \bibinfo {author} {\bibfnamefont {R.~C.}\ \bibnamefont {Leon}}, \bibinfo {author} {\bibfnamefont {C.~H.}\ \bibnamefont {Yang}}, \bibinfo {author} {\bibfnamefont {F.~E.}\ \bibnamefont {Hudson}}, \bibinfo {author} {\bibfnamefont {C.~C.}\ \bibnamefont {Escott}}, \bibinfo {author} {\bibfnamefont {A.}~\bibnamefont {Saraiva}}, \bibinfo {author} {\bibfnamefont {A.~S.}\ \bibnamefont {Dzurak}},\ and\ \bibinfo {author} {\bibfnamefont {A.}~\bibnamefont {Laucht}},\ }\bibfield  {title} {\bibinfo {title} {A high-sensitivity charge sensor for silicon qubits above 1 k},\ }\href@noop {} {\bibfield  {journal} {\bibinfo  {journal} {Nano Letters}\ }\textbf {\bibinfo {volume} {21}},\ \bibinfo {pages} {6328} (\bibinfo {year} {2021})}\BibitemShut {NoStop}%
\bibitem [{\citenamefont {Joecker}\ \emph {et~al.}(2024)\citenamefont {Joecker}, \citenamefont {Stemp}, \citenamefont {Fern{\'a}ndez~de Fuentes}, \citenamefont {Johnson},\ and\ \citenamefont {Morello}}]{joecker2024error}%
  \BibitemOpen
  \bibfield  {author} {\bibinfo {author} {\bibfnamefont {B.}~\bibnamefont {Joecker}}, \bibinfo {author} {\bibfnamefont {H.~G.}\ \bibnamefont {Stemp}}, \bibinfo {author} {\bibfnamefont {I.}~\bibnamefont {Fern{\'a}ndez~de Fuentes}}, \bibinfo {author} {\bibfnamefont {M.~A.}\ \bibnamefont {Johnson}},\ and\ \bibinfo {author} {\bibfnamefont {A.}~\bibnamefont {Morello}},\ }\bibfield  {title} {\bibinfo {title} {Error channels in quantum nondemolition measurements on spin systems},\ }\href@noop {} {\bibfield  {journal} {\bibinfo  {journal} {Physical Review B}\ }\textbf {\bibinfo {volume} {109}},\ \bibinfo {pages} {085302} (\bibinfo {year} {2024})}\BibitemShut {NoStop}%
\bibitem [{\citenamefont {Vaartjes}\ \emph {et~al.}(2025)\citenamefont {Vaartjes}, \citenamefont {Su}, \citenamefont {O'Neill}, \citenamefont {Steinacker}, \citenamefont {Goenka}, \citenamefont {van Blankenstein}, \citenamefont {Yu}, \citenamefont {Wilhelm}, \citenamefont {Jakob}, \citenamefont {Hudson} \emph {et~al.}}]{vaartjes2025maximizing}%
  \BibitemOpen
  \bibfield  {author} {\bibinfo {author} {\bibfnamefont {A.}~\bibnamefont {Vaartjes}}, \bibinfo {author} {\bibfnamefont {R.~Y.}\ \bibnamefont {Su}}, \bibinfo {author} {\bibfnamefont {L.~A.}\ \bibnamefont {O'Neill}}, \bibinfo {author} {\bibfnamefont {P.}~\bibnamefont {Steinacker}}, \bibinfo {author} {\bibfnamefont {G.}~\bibnamefont {Goenka}}, \bibinfo {author} {\bibfnamefont {M.~R.}\ \bibnamefont {van Blankenstein}}, \bibinfo {author} {\bibfnamefont {X.}~\bibnamefont {Yu}}, \bibinfo {author} {\bibfnamefont {B.}~\bibnamefont {Wilhelm}}, \bibinfo {author} {\bibfnamefont {A.~M.}\ \bibnamefont {Jakob}}, \bibinfo {author} {\bibfnamefont {F.~E.}\ \bibnamefont {Hudson}}, \emph {et~al.},\ }\bibfield  {title} {\bibinfo {title} {Maximizing the nondemolition nature of a quantum measurement via an adaptive readout protocol},\ }\href@noop {} {\bibfield  {journal} {\bibinfo  {journal} {arXiv preprint arXiv:2511.10978}\ } (\bibinfo {year} {2025})}\BibitemShut {NoStop}%
\bibitem [{\citenamefont {Yu}\ \emph {et~al.}(2025)\citenamefont {Yu}, \citenamefont {Wilhelm}, \citenamefont {Holmes}, \citenamefont {Vaartjes}, \citenamefont {Schwienbacher}, \citenamefont {Nurizzo}, \citenamefont {Kringh{\o}j}, \citenamefont {Blankenstein}, \citenamefont {Jakob}, \citenamefont {Gupta} \emph {et~al.}}]{yu2025schrodinger}%
  \BibitemOpen
  \bibfield  {author} {\bibinfo {author} {\bibfnamefont {X.}~\bibnamefont {Yu}}, \bibinfo {author} {\bibfnamefont {B.}~\bibnamefont {Wilhelm}}, \bibinfo {author} {\bibfnamefont {D.}~\bibnamefont {Holmes}}, \bibinfo {author} {\bibfnamefont {A.}~\bibnamefont {Vaartjes}}, \bibinfo {author} {\bibfnamefont {D.}~\bibnamefont {Schwienbacher}}, \bibinfo {author} {\bibfnamefont {M.}~\bibnamefont {Nurizzo}}, \bibinfo {author} {\bibfnamefont {A.}~\bibnamefont {Kringh{\o}j}}, \bibinfo {author} {\bibfnamefont {M.~R.~v.}\ \bibnamefont {Blankenstein}}, \bibinfo {author} {\bibfnamefont {A.~M.}\ \bibnamefont {Jakob}}, \bibinfo {author} {\bibfnamefont {P.}~\bibnamefont {Gupta}}, \emph {et~al.},\ }\bibfield  {title} {\bibinfo {title} {Schr{\"o}dinger cat states of a nuclear spin qudit in silicon},\ }\href@noop {} {\bibfield  {journal} {\bibinfo  {journal} {Nature Physics}\ }\textbf {\bibinfo {volume} {21}},\ \bibinfo {pages} {362} (\bibinfo {year} {2025})}\BibitemShut {NoStop}%
\bibitem [{\citenamefont {Nakai}\ \emph {et~al.}(1991)\citenamefont {Nakai}, \citenamefont {Challoner},\ and\ \citenamefont {McDowell}}]{nakai1991forbidden}%
  \BibitemOpen
  \bibfield  {author} {\bibinfo {author} {\bibfnamefont {T.}~\bibnamefont {Nakai}}, \bibinfo {author} {\bibfnamefont {R.}~\bibnamefont {Challoner}},\ and\ \bibinfo {author} {\bibfnamefont {C.~A.}\ \bibnamefont {McDowell}},\ }\bibfield  {title} {\bibinfo {title} {Forbidden nuclear magnetic resonance transitions between singlet and triplet states in spin-12 pair systems in rotating solids},\ }\href@noop {} {\bibfield  {journal} {\bibinfo  {journal} {Chemical Physics Letters}\ }\textbf {\bibinfo {volume} {180}},\ \bibinfo {pages} {13} (\bibinfo {year} {1991})}\BibitemShut {NoStop}%
\end{thebibliography}

\providecommand{\noopsort}[1]{}\providecommand{\singleletter}[1]{#1}%

\end{document}